\documentclass[]{aa}  

\usepackage{natbib}
\bibpunct{(}{)}{;}{a}{}{,}
\usepackage{graphicx}
\usepackage{revsymb}	
\usepackage{amsmath}	
\usepackage[usenames]{color}	
\usepackage{epstopdf}	
\usepackage{txfonts}
\usepackage{amssymb}
\usepackage{color}
\usepackage{geometry} 
\usepackage[parfill]{parskip} 
\usepackage{amssymb}
\usepackage{slantsc}
\usepackage{array}
\usepackage{url}
\usepackage{amsfonts}
\usepackage[colorlinks=true,linkcolor=black, urlcolor=black, citecolor=black]{hyperref}
\usepackage{hyperref} 
\usepackage{sidecap}
\usepackage{graphicx}
\usepackage{xcolor}
\usepackage{xfrac}
\usepackage{subfigure}
\usepackage{epsfig}
\usepackage{threeparttable}
\colorlet{rouge}{red!70!darkgray}

\begin{document}
   \title{Kepler-93: a testbed for detailed seismic modelling and orbital evolution of super-earths around solar-like stars}
\author{J. B\'{e}trisey\inst{1} \and C. Pezzotti\inst{1} \and G. Buldgen\inst{1} \and S. Khan\inst{2,3}  \and P. Eggenberger\inst{1} \and S.~J.~A.~J. Salmon\inst{1} \and A. Miglio\inst{4,5}}
\institute{Observatoire de Genève, Université de Genève, Chemin Pegasi 51, 1290 Versoix, Switzerland\\email: 	\texttt{Jerome.Betrisey@unige.ch}
\and Institute of Physics, Laboratory of Astrophysics, École Polytechnique Fédérale de Lausanne (EPFL), Observatoire de Genève, Chemin Pegasi 51, 1290 Versoix, Switzerland
\and School of Physics and Astronomy, University of Birmingham, Edgbaston, Birmingham B15 2TT, UK
\and Dipartimento di Fisica e Astronomia, Universit{\`a} degli Studi di Bologna, Via Gobetti 93/2, I-40129 Bologna, Italy
\and INAF -- Astrophysics and Space Science Observatory Bologna, Via Gobetti 93/3, I-40129 Bologna, Italy}

\date{\today}

\abstract{The advent of space-based photometry missions such as CoRoT, \textit{Kepler} and TESS has sparkled the rapid development of asteroseismology and its synergies with exoplanetology. In the near future, the advent of PLATO will further strengthen such multi-disciplinary studies. In that respect, testing asteroseismic modelling strategies and their importance for our understanding of planetary systems is crucial.}
{We carried out a detailed modelling of Kepler-93, an exoplanet host star observed by the \textit{Kepler} satellite for which high-quality seismic data is available. This star is particularly interesting as it is a solar-like star very similar to the PLATO benchmark target (G spectral type, $\sim6000K$, $\rm\sim1~M_\odot$ and $\rm\sim1~R_\odot$) and provides a real-life testbed for potential procedures to be used in the PLATO mission.}{We use global and local minimization techniques to carry out the seismic modelling of Kepler-93, varying the physical ingredients of the given theoretical stellar models. We supplement this step by seismic inversion techniques of the mean density. We then use these revised stellar parameters to provide new planetary parameters and simulate the orbital evolution of the system under the effects of tides and atmospheric evaporation.}{We provide the following fundamental parameters for Kepler-93: $\bar{\rho}_\star=1.654\pm0.004$~g/cm$^{3}$, $\rm{M}_\star=0.907\pm0.023 \rm~{M_{\odot}}$, $\rm{R}_\star=0.918\pm0.008 ~R_{\odot}$ and $\rm{Age}_\star=6.78\pm0.32~\rm{Gyr}$. The uncertainties we report for this benchmark star are well within the requirements of the PLATO mission and give confidence in the ability to provide precise and accurate stellar parameters for solar-like exoplanet-host stars. For the exoplanet Kepler-93b, we find $\rm{M_p}=4.01\pm 0.67 \rm~{M}_{\oplus}$, $\rm{R_p}=1.478 \pm 0.014 \rm~{R}_{\oplus}$ and semi-major axis $a=0.0533\pm0.0005$ AU. According to our simulations of the orbital evolution of the system, it seems unlikely that Kepler-93b formed with a mass large enough ($\rm M_{p,initial}>100~M_\oplus$) to be impacted on its orbit by stellar tides.}{For the benchmark case of a solar twin of the PLATO mission, detailed asteroseismic modelling procedures will be able to provide fundamental stellar parameters within the requirements of the PLATO mission. We also illustrate what synergies can be achieved regarding the orbital evolution and atmospheric evaporation of exoplanets once such parameters are obtained. We also note the importance of the high-quality radial velocity follow-up, which here is a limiting factor, to provide precise planetary masses and mean densities to constrain the formation scenarii of exoplanets.}


\maketitle

\section{Introduction}
\label{sec_introduction}
In the last decade, astonishing progresses were achieved in asteroseismology thanks to the revolution initiated by the high quality data from the space based missions CoRoT \citep{2009IAUS..253...71B}, \textit{Kepler} \citep{2010Sci...327..977B} and TESS \citep{2015JATIS...1a4003R}. With the future mission PLATO \citep{2014ExA....38..249R}, the field will again experience a breakthrough as it will observe bright stars in the Southern hemisphere with an expected precision comparable to \textit{Kepler} and will benefit of a better radial velocity follow-up than in the Northern hemisphere. Asteroseismic measurements allow a precise characterisation of stellar parameters, such as mass, radius and age among others, that is hardly reachable with other standard techniques for non-binary stars. In complement to the benefits for asteroseismology, these missions contributed significantly to the development of exoplanets detections \citep{Borde2003, 2010Sci...327..977B, Sullivan2015, Barclay2018, Rauer2018}. In this context, we exploited the asteroseismic data of the \textit{Kepler} Space Telescope, whose capabilities both for exoplanetology and asteroseismology allow great synergies between these fields. Indeed, the exoplanets detection itself does not rely on stellar-model-dependent approaches and thus, a detailed characterization of the host star improves significantly the understanding of the exoplanet(s) \citep[see e.g.][]{2010ApJ...713L.164C,2013ApJ...767..127H,2018ASSP...49.....C}.

We chose to focus on Kepler-93, also known as KOI 69 or KIC 3544595 in the literature. This host star possesses great quality asteroseismic data \citep{2016MNRAS.456.2183D}. In addition, it presents $\log(g)=4.52\pm 0.20$ dex, $[\mathrm{Fe/H}] = -0.18\pm 0.10$ dex, $T_{eff}=5718\pm 100$ K \citep{2018ApJ...861..149F} and $m_V=10.00\pm 0.03$ \citep{2000A&A...355L..27H}, making it a benchmark target for PLATO as the data quality is similar to the expectations for this mission and motivating a new characterisation to reach the precision requirements (2\% for the radius, 15\% for the mass and 10\% for the age). Its stellar parameters were first determined using scaling relations by \citet{2013ApJ...767..127H}. \citet{2014ApJS..210...20M} did a similar analysis and got comparable results. A more detailed modelling of the star, considering the information contained in the individual frequencies, was performed by \citet{2014ApJ...790...12B}, who could provide better constrained parameters. A new analysis using BASTA was conducted by \citet{2015MNRAS.452.2127S}. They estimated the impact of the change of some physical ingredients based on their sample of 33 targets and found stellar parameters with a precision (random+systematic) comparable to \citet{2014ApJ...790...12B}, using a grid of evolutionary models with a fixed Y/Z ratio. Finally, new analyses were performed using a machine learning algorithm \citep{2016ApJ...830...31B,2019A&A...622A.130B}. Their precision is comparable to the previous works, using a single grid with fixed physical ingredients, thus not investigating the systematics coming from their variations.

This star is known to host at least two exoplanets. Thanks to the high photometric precision of the Kepler-93 observations, Kepler-93b was detected during the first four months of \emph{Kepler} data \citep{Borucki2011}. \citet{2014ApJS..210...20M} provided a first estimation of the planetary mass ($\rm 2.6 \pm 2.0~ M_{\oplus}$) on the base of 32 Keck HIRES radial velocity (RV) observations from July 2009 to September 2012. They also detected the presence of a perturbing companion (Kepler-93c), for which they calculated lower limits on the mass ($\rm M > 3~M_{Jup}$) and orbital period ($\rm P > 5$ yr). \citet{2014ApJ...790...12B} derived a very precise best-fit value for the radius measurements of Kepler-93b, being $\rm 1.481 \pm 0.019~R_{\oplus}$, from which they estimated an average planetary density of $6.3 \pm 2.6$ g/cm$^3$, compatible with the one of a rocky-world. On the base of 86 radial velocities observations obtained with the HARPS-N spectrograph on the Telescopio Nazionale Galileo and 32 archival Keck/HIRES observations, \citet{Dressing2015} provided a more precise mass estimate of Kepler-93b ($\rm 4.02 \pm 0.68~M_{\oplus}$) and derived a relatively higher density of $6.88 \pm 1.18$ g/cm$^3$, consistent with a rocky composition primarily of iron and magnesium. The precise determination of the radius and density of Kepler-93b, pointing towards a rocky-world composition with almost no presence of volatile elements, together with the availability of high-quality asteroseismic data of the host star, drive us to perform a study that for the first time relates the rotational history of Kepler-93 with the evolution of the planet. Since Kepler-93b probably lost its atmosphere during the early stages of the evolution, we aim to test the impact of the X-ray and extreme ultraviolet (XUV) fluxes received by the planet accounting for its eventual migration due to the dissipation of tides in the stellar convective envelope, considering different rotational histories of the host star. Specifically, we wish to test whether in the case of a very slow rotator, Kepler 93b could retain a fraction of its primordial atmosphere at its current age. With this kind of study, we aim to understand the mechanisms that concur in shaping the radius valley \citep[see e.g.][]{VanEylen2018}, accounting for a dedicated computation of the high-energy irradiation emitted by the host star along its evolution together with the impact of tides.

In this paper, we provide a detailed modelling of Kepler-93 and then characterise the evolution of Kepler-93b. In Sect.~\ref{sec_forward_modelling}, we carry out a two steps forward modelling, consisting of a global minimization to restrict the parameter space, followed by a local minimisation where we consider extensive changes of the physical ingredients. We also perform a comparison between the direct fit of the individual frequencies and a more refined method, the fit of frequency ratios, to illustrate the relevance of fitting the frequency ratios instead of the individual frequencies. In Sect.~\ref{sec_inversions}, we conduct inversions to provide a more robust estimation of the mean density and test different prescriptions for the surface effects. Based on the improvement of the precision on the stellar parameters, we revise the planetary parameters of Kepler-93b. In Sect.~\ref{sec_Kepler93b}, we study the evolution of Kepler-93b under the impact of stellar tides and evaporation of the planetary atmosphere, by coupling the optimal stellar model of Kepler-93 to our orbital evolution code. Finally, in Sect.~\ref{sec_conclusion}, we draw the conclusions of our study of the Kepler-93 system.

\section{Forward Modelling}
\label{sec_forward_modelling}
The oscillation modes of Kepler-93 were estimated using "peak bagging" by \citet{2016MNRAS.456.2183D}. This technique relies on a  Bayesian approach and on a standard Metropolis-Hastings Monte Carlo Markov Chain (MCMC)\footnote{See e.g. \citet{gelmanbda04} for an introduction to MCMC techniques.}. The output quality is assessed with an unsupervised machine learning Bayesian scheme. In complement to the asteroseismic data, we looked in the literature for spectroscopic measurements that provide good and independent constraints on the stellar metallicity, the effective temperature and the surface gravity. We adopted the values of \citet{2018ApJ...861..149F} with their recommended error\footnote{Note that these data were part of a survey, which explains the conservative precision on the estimated values.}: $\log(g)=4.52\pm 0.20$ dex, $[\mathrm{Fe/H}] = -0.18\pm 0.10$ dex and $T_{eff}=5718\pm 100$ K. Indeed, they provide a reanalysis of the KOI targets with several algorithms, which allowed them to quantify the systematic error associated with the deviations observed between the different routines used to derive the spectroscopic parameters.

The luminosity was computed using the following formula:
\begin{equation}
\log\left(\frac{L}{L_\odot}\right) = -0.4\left(m_\lambda + BC_\lambda -5\log d + 5 - A_\lambda -M_{\mathrm{bol},\odot}\right) \, ,
\end{equation} 
where $m_\lambda$, $BC_\lambda$, and $A_\lambda$ are the magnitude, the bolometric correction, and the extinction in a given band $\lambda$. We used the 2MASS $K_s$-band magnitude properties. The bolometric correction was estimated using the code written by \citet{2014MNRAS.444..392C,2018MNRAS.475.5023C}, and the extinction was inferred with the \citet{2018MNRAS.478..651G} dust map. A value of $M_{\mathrm{bol},\odot} = 4.75$ was adopted for the solar bolometric magnitude. Two different methods have been tested to obtain the distance $d$ in pc from \textit{Gaia} \citep{2018A&A...616A...1G}: either directly inverting the parallax or using the published distance from \citet{2018AJ....156...58B}. Both lead to similar luminosity estimates, $L = 0.82\pm 0.03~L_\odot$, which is the value adopted in this paper.

All the stellar evolutionary sequences in this work were computed with the Code Li{\'e}geois d'{\'E}volution Stellaire (CLES) \citep{2008Ap&SS.316...83S} and the adiabatic frequencies and eigenfunctions with the Li{\`e}ge Oscillation Code (LOSC) \citep{2008Ap&SS.316..149S}. The modelling is articulated in two steps. In a first step, a global minimization is conducted to restrict the parameter space. Here, we used the AIMS software \citep{2019MNRAS.484..771R} which is a MCMC based algorithm. In a second time, the impact of the physical ingredients was investigated. For this purpose, we used a local minimization method, the Levenberg-Marquardt algorithm \citep[see e.g.][]{SamRoweisLevenbergMarquardt}. The MCMC algorithm runs with 4 free parameters, the mass $M$, the age $\tau$ and the initial chemical composition, with $X_0$ the initial hydrogen mass fraction and $Z_0$ the heavy elements mass fraction. For the Levenberg-Marquardt, we included one additional free parameter, the mixing-length parameter $\alpha_{\mathrm{MLT}}$. This technique is indeed much faster and allows a more detailed exploration of the parameter space, but only at a local level. If surface effects are considered, they are described by two additional free parameters.

\subsection{Global Minimization}
\label{sec_global_minimization}
For the global minimization, a grid of models was first generated by varying the mass ($M_{min}=0.70M_\odot$, $M_{max}=1.10M_\odot$, $M_{step}=0.02M_\odot$), the initial hydrogen mass fraction ($X_{0,min}=0.72$, $X_{0,max}=0.77$, $X_{0,step}=0.01$) and the initial heavy elements mass fraction ($Z_{0,min}=0.006$, $Z_{0,max}=0.013$, $Z_{0,step}=0.001$) and keeping the following physical ingredients fixed. We used the AGSS09 abundances \citep{2009ARA&A..47..481A} and the OPAL opacities \citep{1996ApJ...464..943I}, supplemented by the \citet{2005ApJ...623..585F} opacities at low temperature and the electron conductivity by \citet{1999A&A...346..345P} and \citet{2007ApJ...661.1094C}. The FreeEOS equation of state \citep{2012ascl.soft11002I} is used and the microscopic diffusion is described by the formalism of \citet{1994ApJ...421..828T} but with the screening coefficients of \citet{1986ApJS...61..177P}. Convection is implemented using the mixing-length theory (MLT) as in \citet{1968pss..book.....C} and the nuclear reaction rates are taken from \citet{2011RvMP...83..195A}. The mixing-length parameter $\alpha_{\mathrm{MLT}}$ is fixed at a solar-calibrated value. Finally, for the atmosphere modelling, we use the $T(\tau)$ relation described by Model-C in \citet{1981ApJS...45..635V} (here after VAL-C).

The minimization is performed with the AIMS software \citep{2019MNRAS.484..771R}, which is based on a MCMC algorithm \citep[\textsc{emcee},][]{2013PASP..125..306F} and a Bayesian statistics approach to provide the probability distributions for the stellar parameters. The MCMC uses an interpolation scheme to sample between the grid points. This interpolation reduces the finesse of the grid required to have a sufficient sampling of the parameter space. Note that this approach is well suited for main sequence stars whose solution is not too degenerate. We remark that in our case, the combination of small steps for the grid and the interpolation within these steps results in a very fine exploration of the parameter space. All the priors of the free variables are uninformative, except the age prior which is the uniform distribution on the interval $[0,14]$ Gyr. We use Gaussian priors for the constraints. In addition to the classical constraints  (in our case the effective temperature, luminosity and metallicity), we consider two sets of seismic constraints in AIMS, the observed individual frequencies and the frequency ratios. We used the $r_{01}$ and $r_{02}$ ratios, according to the definitions of \citet{2003A&A...411..215R}. It is possible to replace the $r_{01}$ ratios by the $r_{10}$, but since they are not independent, they should not be used together, as it will introduce a bias \citep{2018arXiv180807556R}. In general, the fit of the ratios requires an estimate of the mean density as additional constraint since the information about the mean density is suppressed from the ratios in their definition, as they are normalized by the large frequency separation that is a proxy of the mean density \citep{1967AZh....44..786V}. We therefore carried out a mean density inversion\footnote{Please refer to Sect. \ref{sec_inversions} for a detailed explanation of what is a mean density inversion.} on the model resulting from the fit of the frequencies. Since this inverted mean density is the result of the inversion of only one model, we considered a conservative error. The final mean density quoted in Table \ref{tab_revised_stellar_parameters} is the result of a more elaborated procedure, also described in Sect. \ref{sec_inversions}. Hence, we adopted a mean density estimate of $1.654\pm 0.010$ g/cm$^3$. Note that this guess is consistent with the final mean density provided in this work.

The first MCMC minimization was conducted with the individual frequencies as constraints and the two-terms surface effects correction from \citet{2014A&A...568A.123B} (cf. Eqs. \eqref{eq_BG2014_1} and \eqref{eq_BG2014_2}). The best MCMC model was coherent, as it reproduced the observed échelle diagram. However, the mode identification of AIMS revealed a shift with respect to the identification of \citet{2016MNRAS.456.2183D}. Indeed, as shown in Appendix \ref{sec_appendix_mode_identification}, the asymptotic behaviour of the $\epsilon_{nl}$ phases \citep{2012ApJ...751L..36W,2016A&A...585A..63R} with the identification of \citet{2016MNRAS.456.2183D} is inconsistent, as they lie significantly below 1, what is unexpected. The corrected identification can be found in Appendix \ref{sec_appendix_observational_data}. After the mean density inversion on the model resulting from the fit of the frequencies, we performed a second MCMC minimization where we added the inverted mean density to the constraints and replaced the individual frequencies by the frequency ratios. Note that the fit of the ratios was done without surface effects. The ratios are indeed designed to reduce the impact of surface effects \citep{2003A&A...411..215R} and it is thus difficult to estimate them with this set of constraints. The corner plot of the MCMC fitting the frequency ratios can be found in Appendix \ref{sec_appendix_mcmc_corner_plot}. This plot shows the distributions and correlations of the stellar mass, helium content, metallicity, age (that are the optimized parameters) and radius. The addition of the inverted mean density in the constraints especially improves the precision on the determination of the mass and the radius. The mean density of the forward modelling is strongly constrained by the assumption on the value and uncertainty on the inverted mean density, which is not an issue since a mean density inversion is quasi-model independent and we also assumed a conservative uncertainty to avoid biases. For Kepler-93, we found that the $r_{01}$ ratios improve the precision on the age but the mean density does not. Indeed, the central hydrogen value $X_c$ (and hence the age) is already well constrained by the seismic information in the case of Kepler-93. For stars with a less well constrained $X_c$, it would be interesting to study if the gain on the mass precision thanks to the mean density could help to better constrain the age.

\begin{figure}
\centering
\includegraphics[scale=0.32]{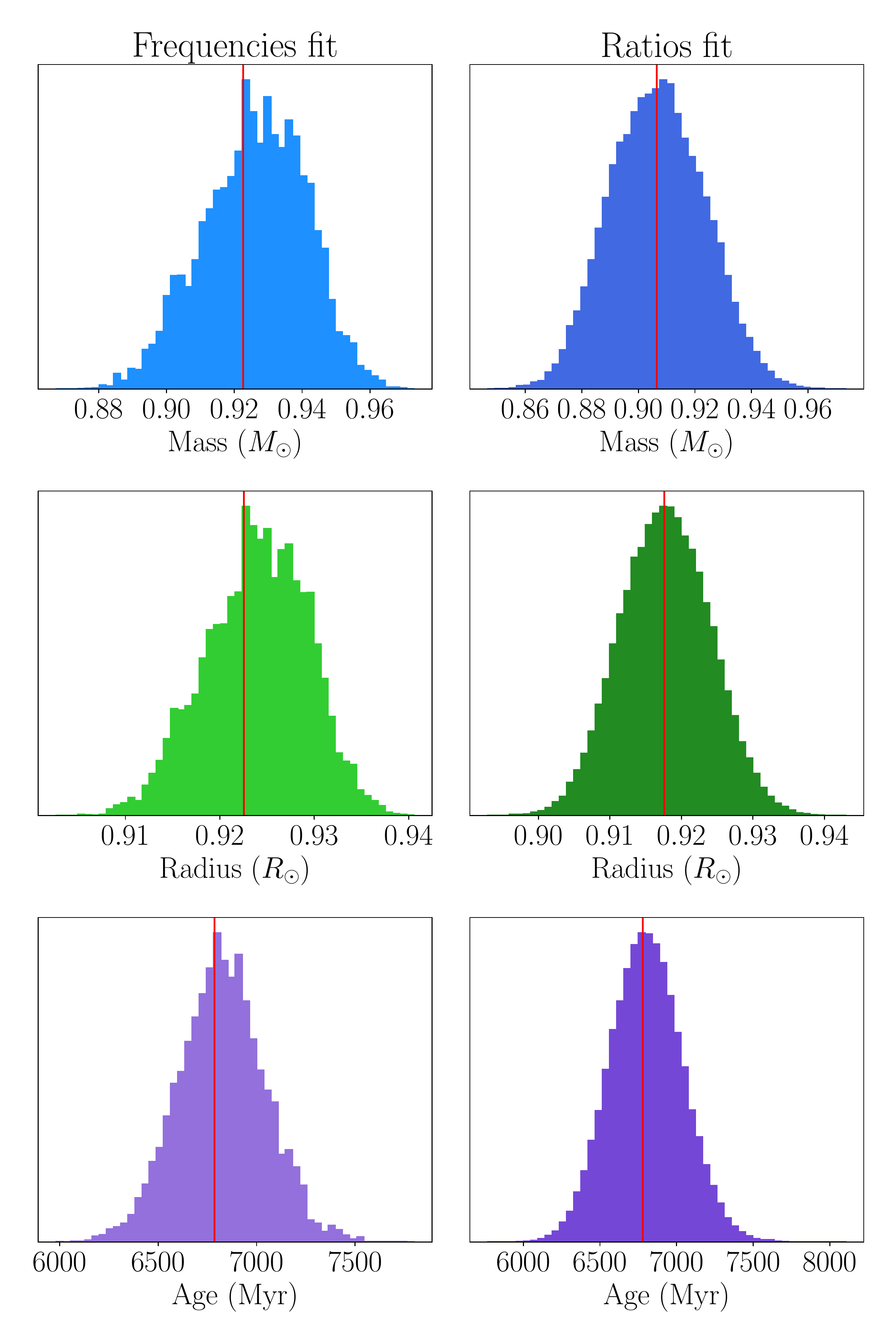} 
\caption{Comparison of the histograms between the fit of the individual frequencies (left column) and of the frequency ratios $r_{01}$ \& $r_{02}$ (right column). \textit{Upper panel:} Histogram for the stellar mass. \textit{Middle panel:} Histogram for the stellar radius. \textit{Lower panel:} Histogram for the stellar age. The red vertical lines indicate the optimal estimates predicted by the MCMC.}
\label{fig_AIMS}
\end{figure}

The main stellar parameters for the two sets of constraints are displayed in Table \ref{tab_AIMS} and a comparison between the histograms of both sets is shown in Fig. \ref{fig_AIMS}. On one hand, the fit of the frequencies is more sensitive to the grid. Indeed, small peaks can be seen on the top of the histograms, which indicates that the walkers tend to get stuck on the grid points during the exploration of the parameter space. This effect is not observed for the fit of the ratios and their smooth histograms. On the other hand, the fit of the frequencies tends to overestimate the stellar mass. This trend is linked with the treatment of the surface effects that is empirical and thus not optimal. This overestimation is problematic since \citet{2019MNRAS.484..771R} found that their software produces very peaked distributions when fitting the frequencies, what can lead to a precise but biased mass value. A similar behaviour was also observed by \citet{2005A&A...441..615M}, \citet{2019A&A...630A.126B} and \citet{2021A&A...646A...7S}.

\begin{figure}
\centering
\includegraphics[scale=0.45]{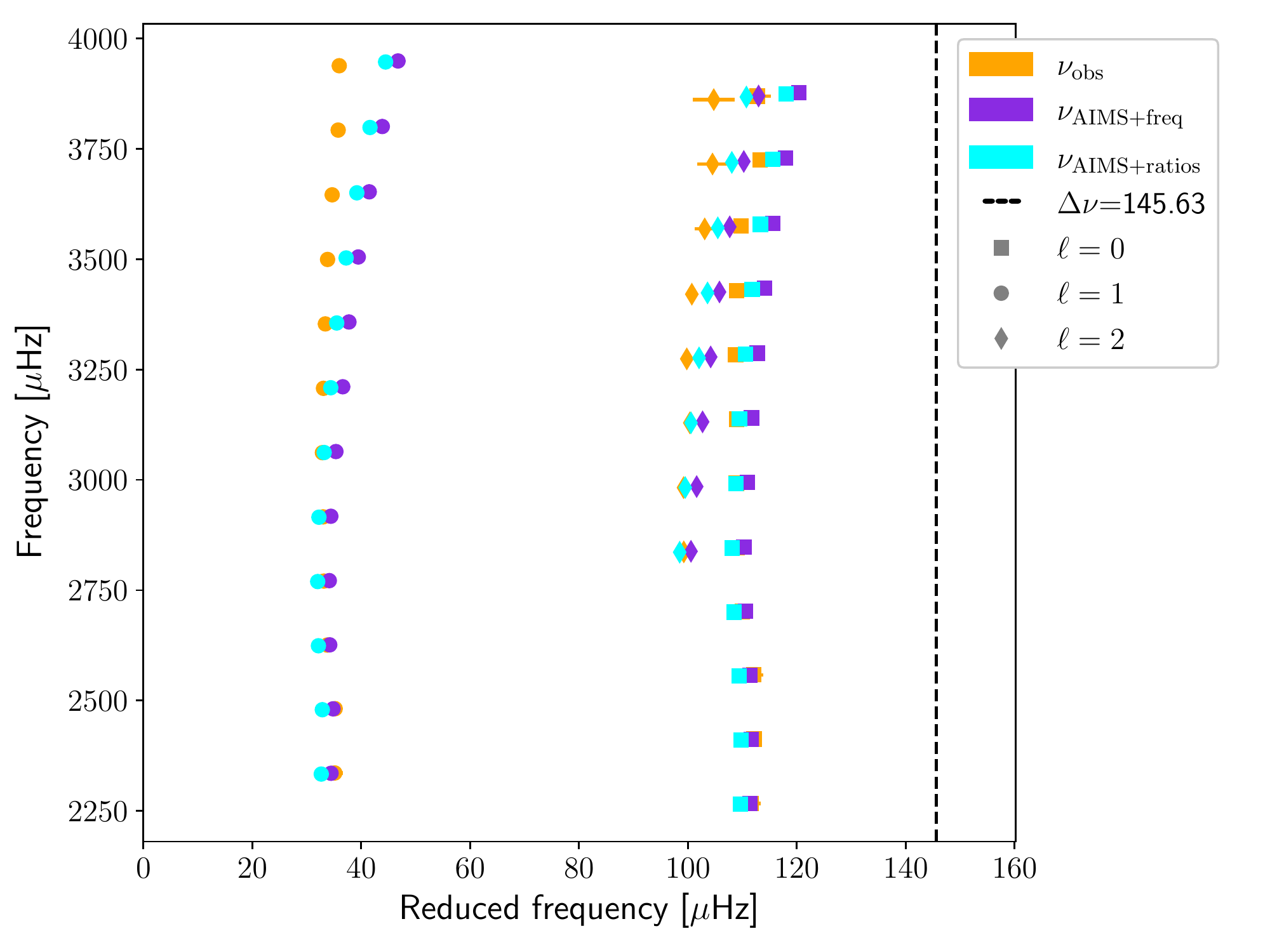}
\caption{Echelle diagram of the AIMS results for the fit of the individual frequencies (purple) and of the frequency ratios $r_{01}$ \& $r_{02}$ (cyan). The modelled frequencies are displayed without a correction for the surface effects. The large separation (black dashed line) is computed with the observed frequencies (orange) and with the definition of \citet{2012A&A...539A..63R}. The squares are radial modes, the circles are dipole modes and the diamonds are quadrupole modes.}
\label{fig_echelle_AIMS}
\end{figure}

\begin{figure}
\centering
\includegraphics[scale=0.48]{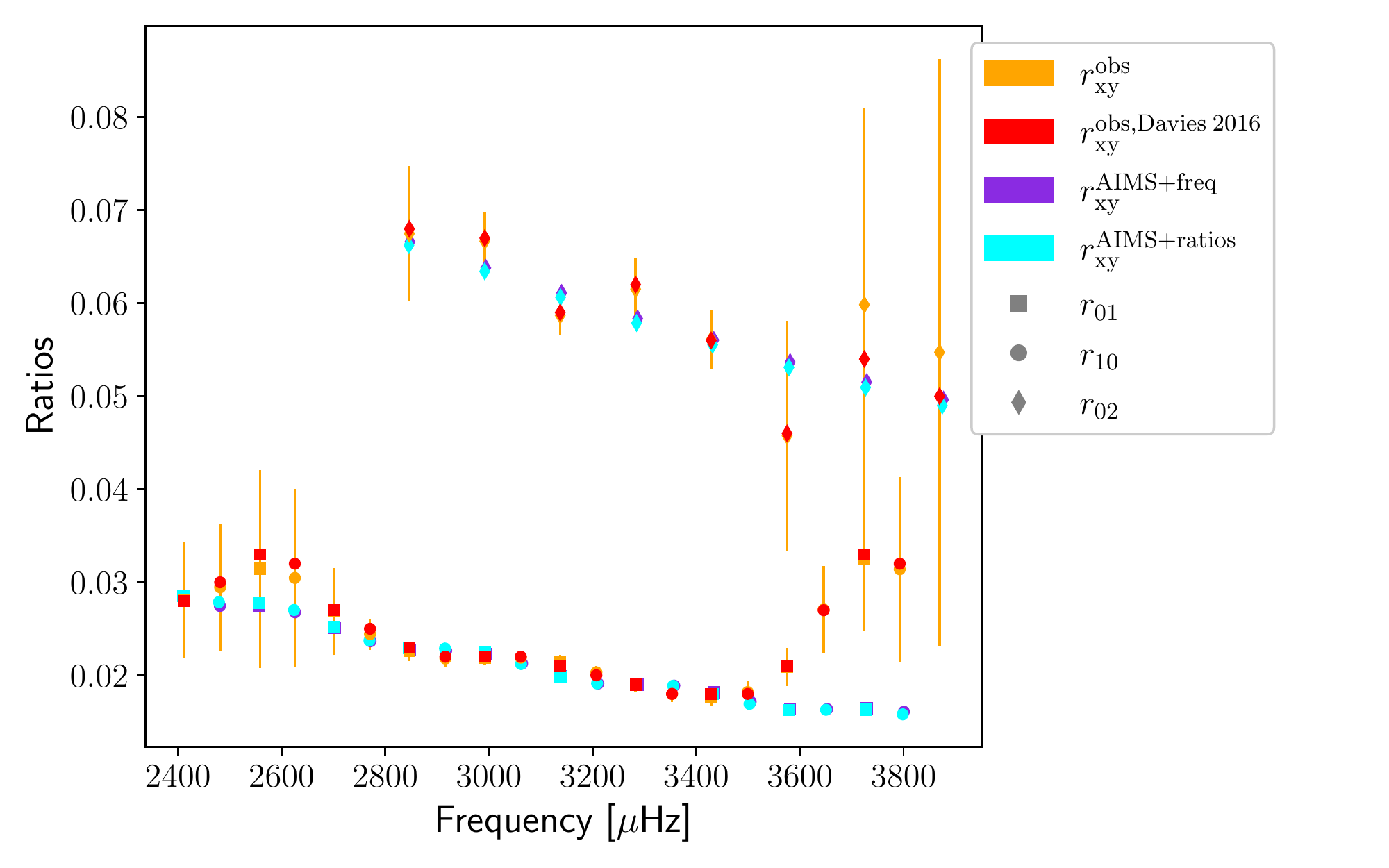}
\caption{Observed and modelled frequency ratios $r_{01}$, $r_{10}$ and $r_{02}$ of Kepler-93. In red, the ratios computed by \citet{2016MNRAS.456.2183D} without errorbar for readability reasons. Remark that these errors are similar to the ones found in this work. The observed ratios of our study (orange) are computed with the definitions of \citet{2003A&A...411..215R}, while the ratios from \citet{2016MNRAS.456.2183D} come from its fitting procedure. The modelled ratios are plotted in purple for the fit of the frequencies and in cyan for the fit of the ratios. The bump on the right of the plot for the $r_{01}$ and $r_{10}$ ratios is likely due to surface activity during the measurement.}
\label{fig_ratios_AIMS}
\end{figure}

In Fig. \ref{fig_echelle_AIMS}, we show the échelle diagram of the results from the fit of both sets of seismic constraints. They are both in agreement with the observed frequencies. The ratios $r_{01}$, $r_{10}$ and $r_{02}$, displayed in Fig. \ref{fig_ratios_AIMS} are also well reproduced. The bump at high frequencies for the $r_{01}$ and $r_{10}$ is likely due to surface activity. Indeed, a shift on the frequencies due to surface activity was observed for the Sun by \citet{2018MNRAS.480L..79H,2020MNRAS.493L..49H} or in the \emph{Kepler} data by \citet{2018ApJS..237...17S}. Concerning the ratios, \citet{2021MNRAS.502.5808T} observed that surface activity has an impact on the $r_{02}$ ratios, especially at high frequencies. By now, no investigation was conducted on the $r_{01}$ and $r_{10}$, but based on the literature and since the bump is at high frequencies, it is very possible that this is the case.

The initial helium mass fraction $Y_{0}$ of the model fitting the frequencies is slightly lower than the value of the primordial helium mass fraction $Y_p\simeq 0.247$ \citep{2018PhR...754....1P}. Even though Kepler-93 is a metal-poor star, we do not expect a helium mass fraction close to the primordial value. However, the precision on the frequencies is insufficient to make use of glitch fitting techniques to directly constrain the helium abundance in the envelope. Hence, $Y_0$ is mainly indirectly constrained by the non-seismic observables (i.e. the metallicity, the effective temperature and the luminosity) which are limited and have a conservative precision as they come from a survey. Consequently, $Y_0$ is weakly constrained and has a large uncertainty, as shown in Appendix \ref{sec_appendix_mcmc_corner_plot}. We still note that the primordial helium mass fraction is contained within the first sigma and that the observed [Fe/H] has also a large uncertainty. The value of $Y_0$ for the fit of the ratios is slightly bigger than $Y_p$ but the same argument applies. Therefore, the large uncertainty on the helium mass fraction just indicates that it is not possible to precisely contrain this quantity with the current data. We remark that Kepler-93 and low metallicity targets more generally are interesting in this aspect as excellent spectroscopic and seismic data would allow to point out eventual inconsistencies in the stellar models based on the predicted initial helium mass fraction.

\begin{table}
\centering
\caption{Main stellar parameters predicted by AIMS for the models based on the fit of the frequencies and on the fit of the $r_{01}$ \& $r_{02}$ ratios. The errors are statistical errors.}
\begin{tabular}{ccc}
\hline 
Seismic constraints & $\nu_{nl}$ & $r_{01}$ \& $r_{02}$ \\ 
\hline \hline
$M/M_\odot$ & $0.923\pm 0.024$ & $0.907\pm 0.017$ \\ 
$R/R_\odot$ & $0.923\pm 0.017$ & $0.918\pm 0.006$ \\ 
$\tau$ (Myr) & $6786\pm 259$ & $6780\pm 235$ \\ 
$\bar{\rho}$ (g/cm$^3$) & $1.655\pm 0.057$ & $1.653\pm 0.010$ \\
$X_0$ & $0.753\pm 0.014$ & $0.744\pm 0.014$ \\ 
$Y_0$ & $0.237\pm 0.014$ & $0.247\pm 0.014$ \\ 
$Z_0$ & $(9.8\pm 1.3)\cdot 10^{-3}$ & $(9.1\pm 1.2)\cdot 10^{-3}$ \\ 
\hline 
\end{tabular}
\label{tab_AIMS}
\end{table}

The goal of the global minimization was to restrict the parameter space to start varying the physical ingredients and explore locally the parameter space in a more detailed way. We will also be able to test the impact of different prescriptions to estimate the surface effects.

\subsection{Local Minimization and Impact of the physical Ingredients}
\label{sec_local_minimization}
\begin{figure}[h]
\centering
\includegraphics[scale=0.23]{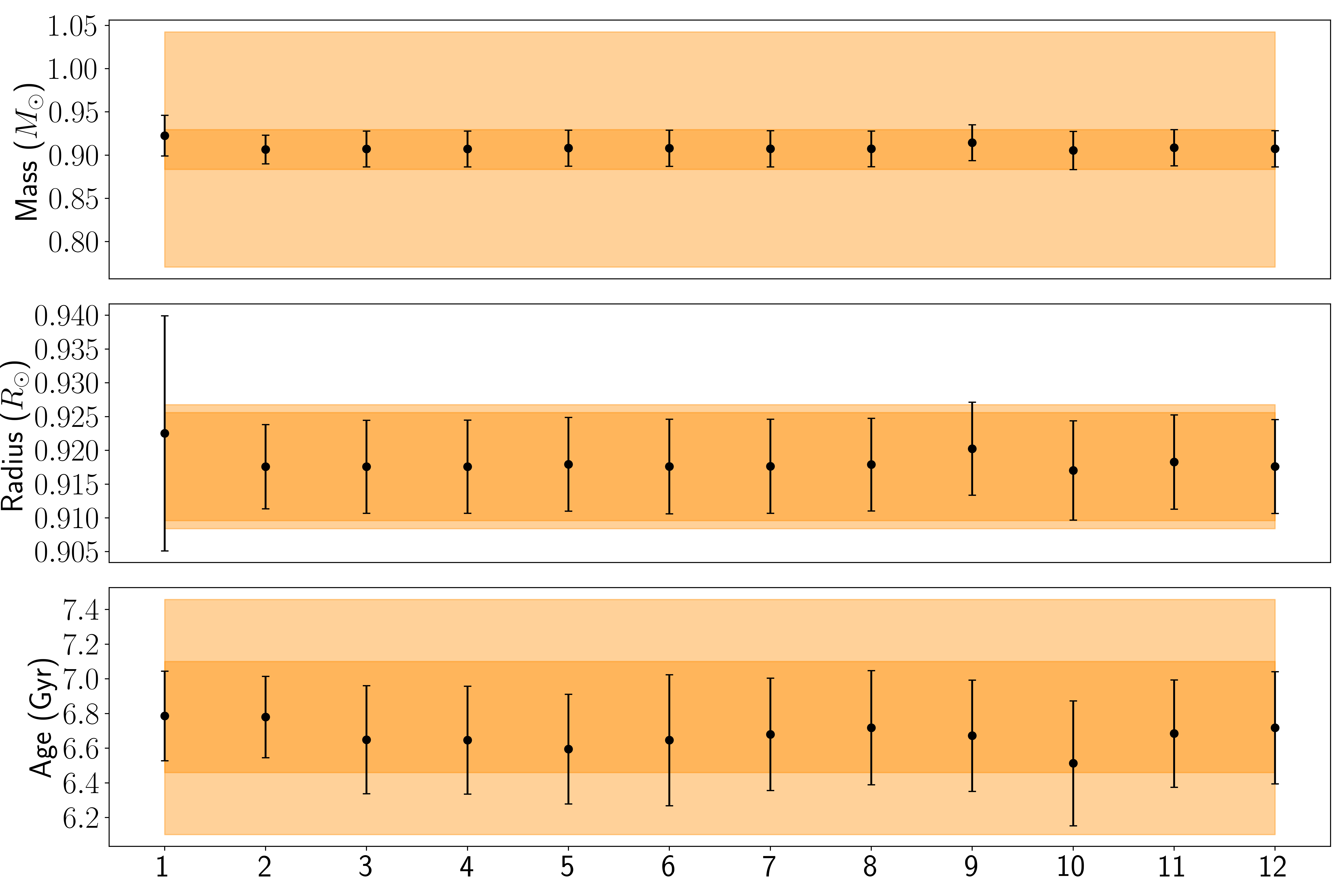}
\caption{Graphical visualization of Table \ref{tab_parameters_levenberg}. The numbers $i$ of the x-axis should be read as \textit{Model $i$}. From top to bottom are displayed the mass, radius and age of the different models. The dark orange band indicates the final one sigma interval (cf. Table \ref{tab_revised_stellar_parameters}) and the light orange band the precision requirements of PLATO (15\% in mass, 1-2\% in radius and 10\% in age). The optimal stellar parameters are well within the PLATO precision requirements.}
\label{fig_visualization_table2}
\end{figure}

In this section, we tested a wide range of prescriptions for the physical processes acting in the star, focussing first on the impact of the abundances. For this purpose, we tested the GN93 \citep{1993oee..conf...15G}, the AGSS09 \citep{2009ARA&A..47..481A} and the AGSS09Ne \citep[AGSS09 with the Neon revised according to][]{2015ApJ...800..110L,2018ApJ...855...15Y} abundances. Then, we investigated the impact of the opacities, by considering the OPAL \citep{1996ApJ...464..943I} and the OPLIB \citep{2016ApJ...817..116C} opacities. The impact of a turbulent diffusion (DT) \citep{1991ApJ...380..238P} was also tested. The turbulent coefficient has the form:
\begin{align}
D_{\mathrm{turb}} = D\left(\frac{\rho_{\mathrm{bcz}}}{\rho(r)}\right)^n,
\end{align}
where $D$ \& $n$ are free parameters, $\rho_{\mathrm{bcz}}$ is the density at the base of the convective zone and $\rho(r)$ the density profile. We considered two sets of coefficients, respectively from \citet{2017MNRAS.472L..70B} and \citet{1991ApJ...380..238P}. The first set is given by $D_1=50$ \& $n_1=2$ and the second one by $D_2=7500$ \& $n_2=3$. The mixing-length parameter $\alpha_{\mathrm{MLT}}$ is one of the free variables. The rest of the physical ingredients are the same as for the global minimization. A summary of the models used in this work can be found in Table \ref{tab_parameters_levenberg}. A graphical representation of Table \ref{tab_parameters_levenberg} is displayed in Fig. \ref{fig_visualization_table2}, with the PLATO precision requirements (light orange band) and the final one sigma interval (cf. Table \ref{tab_revised_stellar_parameters}) of the optimal stellar parameters of Kepler-93 (dark orange band).

\begin{table*}
\centering
\caption{Physical ingredients of the different models considered in this work. The errors are statistical errors. The Model$_1$ is based on the fit of the individual frequencies with AIMS and the Model$_2$ on the fit of the frequency ratios. The rest of the models are the results of the Levenberg minimization. The models 3-6 have respectively an overshoot $\alpha_{\mathrm{ov}}=0.05,0.10,0.15,0.20$.}
\resizebox{\linewidth}{!}{
\begin{tabular}{cccccccccc}
\hline 
\textbf{Name} & \textbf{Mass} (M$_\odot)$ & \textbf{Radius} (R$_\odot)$ & \textbf{Age} (Gyr) & $X_0$ & $(Z/X)_0$ & \textbf{Opacities} & \textbf{Abundances} & \textbf{Diffusion} & \textbf{Convection} \\ 
\hline \hline 
Model$_{1}$ & $0.923\pm 0.024$ & $0.923\pm 0.017$ & $6.79\pm 0.26$ & $0.753\pm 0.14$ & $0.0130\pm 0.0017$ & OPAL & AGSS09 & Paquette & MLT \\ 
Model$_{2}$ & $0.907\pm 0.017$ & $0.918\pm 0.006$ & $6.78\pm 0.23$ & $0.744\pm 0.14$ & $0.0123\pm 0.0017$ & OPAL & AGSS09 & Paquette & MLT \\ 
Model$_{3}$ & $0.907\pm 0.021$ & $0.918\pm 0.007$ & $6.65\pm 0.31$ & $0.744\pm 0.17$ &  $0.0123\pm 0.0026$ & OPAL & AGSS09 & Paquette & MLT + OV1 \\ 
Model$_{4}$ & $0.907\pm 0.021$ & $0.918\pm 0.007$ & $6.65\pm 0.31$ & $0.744\pm 0.17$ &  $0.0123\pm 0.0026$ & OPAL & AGSS09 & Paquette & MLT + OV2 \\ 
Model$_{5}$ & $0.908\pm 0.021$ & $0.918\pm 0.007$ & $6.60\pm 0.32$ & $0.743\pm 0.17$ &  $0.0123\pm 0.0027$ & OPAL & AGSS09 & Paquette & MLT + OV3 \\ 
Model$_{6}$ & $0.908\pm 0.021$ & $0.918\pm 0.007$ & $6.65\pm 0.38$ & $0.743\pm 0.17$ &  $0.0123\pm 0.0026$ & OPAL & AGSS09 & Paquette & MLT + OV4 \\ 
Model$_{7}$ & $0.907\pm 0.021$ & $0.918\pm 0.007$ & $6.68\pm 0.32$ & $0.744\pm 0.17$ & $0.0122\pm 0.0027$ & OPAL & AGSS09 & Paquette + DT1 & MLT \\ 
Model$_{8}$ & $0.907\pm 0.021$ & $0.918\pm 0.007$ & $6.72\pm 0.33$ & $0.744\pm 0.17$ & $0.0122\pm 0.0027$ & OPAL & AGSS09 & Paquette + DT2 & MLT \\ 
Model$_{9}$ & $0.914\pm 0.021$ & $0.920\pm 0.007$ & $6.67\pm 0.32$ & $0.743\pm 0.16$ & $0.0149\pm 0.0032$ & OPAL & GN93 & Paquette & MLT \\ 
Model$_{10}$ & $0.905\pm 0.022$ & $0.917\pm 0.007$ & $6.51\pm 0.36$ & $0.745\pm 0.18$ & $0.0127\pm 0.0027$ & OPLIB & AGSS09 & Paquette & MLT \\ 
Model$_{11}$ & $0.909\pm 0.021$ & $0.918\pm 0.007$ & $6.68\pm 0.31$ & $0.743\pm 0.17$ & $0.0125\pm 0.0027$ & OPAL & AGSS09Ne & Paquette & MLT \\ 
Model$_{12}$ & $0.907\pm 0.021$ & $0.918\pm 0.007$ & $6.72\pm 0.32$ & $0.743\pm 0.17$ & $0.0123\pm 0.0026$ & OPAL & AGSS09 & Paquette+PartIon & MLT \\
\hline
\end{tabular}
}
\label{tab_parameters_levenberg}
\end{table*}

\begin{figure*}
\centering
\includegraphics[scale=0.50]{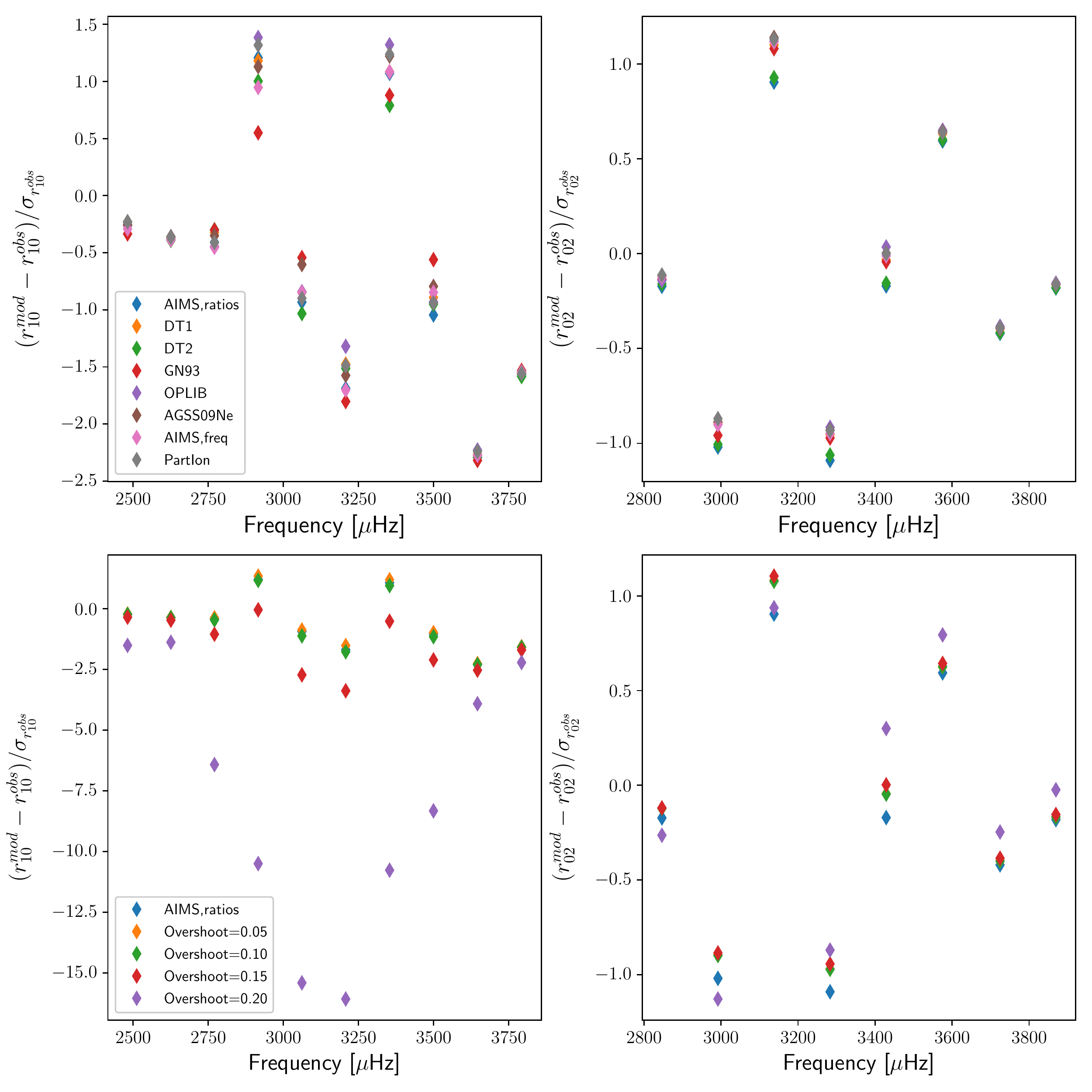}
\caption{\textit{Upper panel:} Comparison of the $r_{10}$ (left) and $r_{02}$ (right) frequency ratios obtained by varying the physical ingredients and without overshoot. The model with the GN93 abundances slightly stand out. \textit{Lower panel:}  Comparison of the $r_{10}$ (left) and $r_{02}$ (right) frequency ratios obtained by varying the overshoot parameter. The model labelled with "AIMS,ratios" is without overshoot. On the lower left panel, there is no visible difference between the blue and orange symbols. On the lower right panel, orange and green symbols are also largely superimposed.}
\label{fig_ratios_comp}
\end{figure*}

The Levenberg-Marquardt algorithm is known to have difficulties to properly estimate the uncertainty. Indeed, it finds the minimum and only then estimates the errors based on the steps that conducted to this optimal solution. Two problems arise with this procedure. First, the minimum can be local and the errors will be underestimated in this case and second, if the steps get too small, the inversion of the Hessian matrix may fail and output overestimated uncertainties. In order to avoid these problems, we estimated with AIMS and its robust error estimation, the uncertainties expected with the same set of constraints and adapted our steps to reproduce them. We did not use the $r_{01}$ (neither the $r_{10}$) as constraints in the Levenberg-Marquardt minimization. Indeed the sole fit of the $r_{02}$, the inverted mean density and the non-seismic constraints is able to well reproduce all the ratios, as shown in Fig. \ref{fig_ratios_comp}. In addition, their inclusion would make the algorithm unstable. The error on the radius is not provided by the Levenberg as it is not one of the optimized variables, but it can be estimated using a rule of thumb. Indeed, the relative uncertainty on the radius is about one third of the one of the mass, as a result of the inclusion of the inverted mean density in the constraints. This was observed by \citet{2019A&A...630A.126B} for Kepler-444 and the same behaviour also appears with Kepler-93 looking at the AIMS results. We stress out that the errors provided by the Levenberg-Marquardt miminization are not considered as robust. However, this is not a problem since we only used the results from the local minimization to estimate the systematics due to the physical ingredients and our final value are based on the model from AIMS whose error estimation procedure is far more reliable and robust. Also note that the $\alpha_{\mathrm{MLT}}$ parameter, let as a free variable in the minimization process and without constraining prior interval, stayed close to the solar-calibrated value\footnote{See Sect. \ref{sec_global_minimization}, 1$^{\rm{st}}$ paragraph for the associated set of physical ingredients and remark that we used a VAL-C atmosphere.} $\alpha_{\mathrm{MLT},\odot}\simeq 2.05$. Indeed, the estimated uncertainty for this quantity is about 0.1 and the standard deviation of the sample is 0.02 what is well below one sigma. The fact that $\alpha_{\mathrm{MLT}}$ stays close to the solar-calibrated value indicates that we cannot constrain this quantity with our data.

A comparison of the $r_{10}$ and $r_{02}$ frequency ratios obtained with the different models of this work without overshoot is shown in the upper panel of Fig. \ref{fig_ratios_comp}. These results are relatively close to each other, indicating that a change of the physical ingredients does not have a significant impact. We still remark that the model with the GN93 abundances slightly stands out. With theses abundances, we have a change in the initial chemical composition, which modifies the behaviour of the sound speed in the radiative zone. This will change the modelled stellar radius and consequently the ratios. The reader is invited to consider the Fig. \ref{fig_ratios_comp} with caution. We displayed the quantity $(r_{xy}^{\mathrm{mod}}-r_{xy}^{\mathrm{obs}})/\sigma_{xy}^{\mathrm{obs}}$ which boosts the differences between the observations and the modelled ratios, especially if the precision on the observed ratio is very high. But, as it is illustrated in Fig. \ref{fig_ratios_AIMS}, the models without overshoot reproduced well the observed ratios. 

\begin{figure*}
\centering
\includegraphics[scale=0.5]{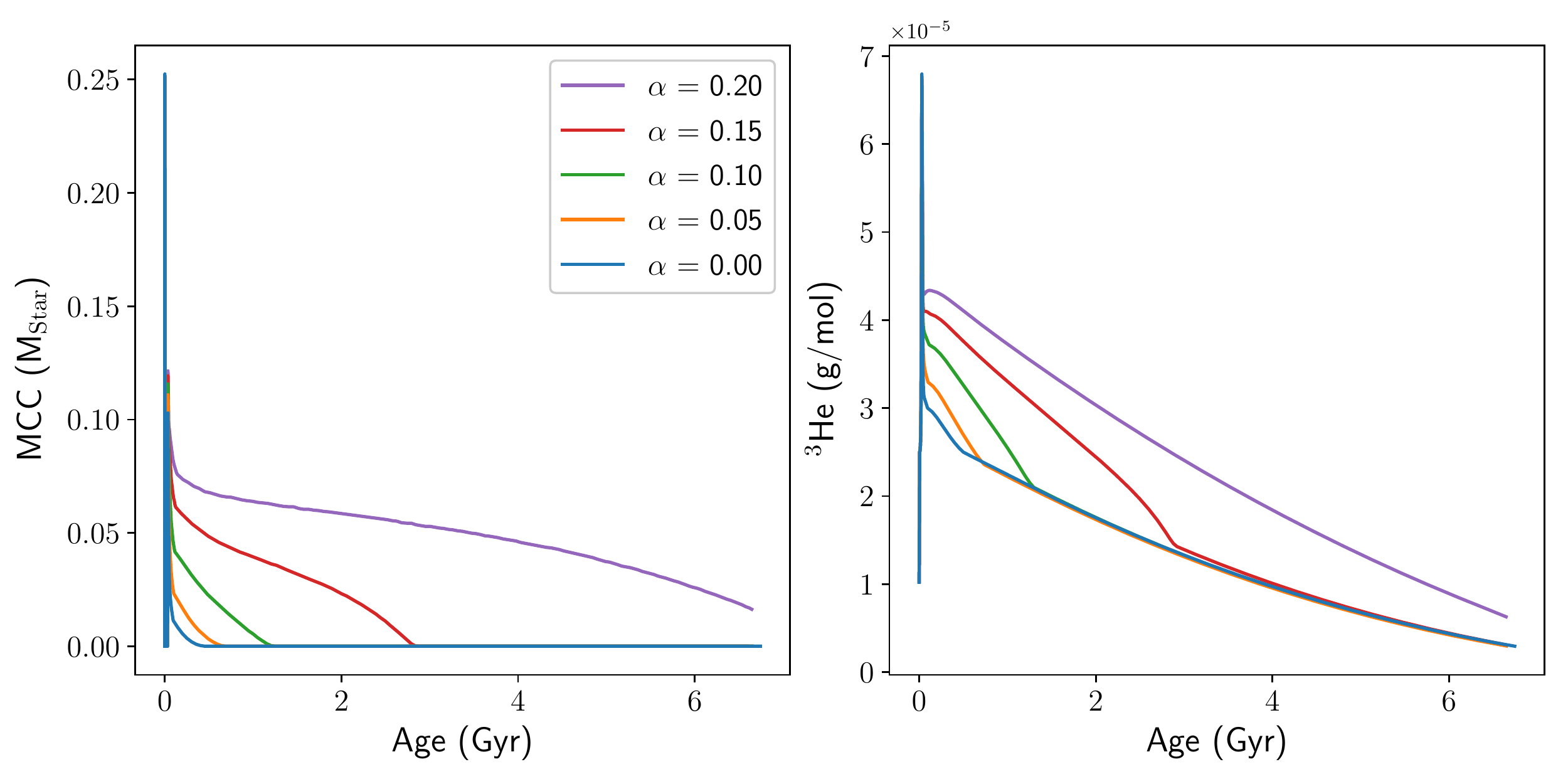}
\caption{\textit{Left:} Comparison of the evolution of the mass of the convective core (MCC) with different overshooting efficiencies. Note that the lilac model with a strong overshoot does not reproduce the observed frequency ratios. Hence, a convective core is not expected to survive until present time. \textit{Right:} Comparison of the evolution of central $^3$He abundance with different overshoots.}
\label{fig_MCC}
\end{figure*}

For Kepler-444 and HD 203608, who are targets with mass and metallicity similar to Kepler-93, it was found that a transitory convective formed during their evolution and its trace is still visible in the observed ratios. Indeed, in the case of Kepler-444, the survival of the convective core during a significant part of the stellar lifetime showed an impact on the observed frequencies and ratios \citep{2019A&A...630A.126B}. \citet{2010A&A...514A..31D} found strong evidence that the convective core of HD 203608 survived until present times. Therefore, we investigated the survival of the convective core of Kepler-93 to understand if this effect should be taken into account in the modelling. The evolution and lifetime of a convective core is linked to the $^3$He/H ratio. More precisely, the critical point is the temperature sensitivity of the out-of-equilibrium $^3$He burning, which concentrates the energy generation in a smaller region and hence increases the radiative gradient. As illustrated in the right panel of Fig. \ref{fig_MCC}, without overshooting, this ratio reaches quickly an equilibrium value and consequently the ppI chain reaches equilibrium. In this case, the radiation is sufficient to evacuate the energy flux. With the consideration of an overshoot, which brings more $^3$He in the central regions, its out-of-equilibrium burning will last longer. Since the out-of-equilibrium $^3$He burning temperature is too high for radiation to evacuate the heat, the lifetime of the convective core will be extended and it can remain during the main sequence \citep{1985SoPh..100...21R}. In the lower panel of Fig. \ref{fig_ratios_comp}, we tested several overshoot regimes. The models with low overshoot do not depart from the models with none, while the models with strong overshoot do not reproduce the observed ratios. The lifetime of the convective core for Kepler-93 is plotted in the left panel of Fig. \ref{fig_MCC}. Without overshoot, the convective core almost instantaneously disappears. Since the model with strong overshoot does not reproduce the observed ratios, the lifetime of a convective core is not expected to exceed 3 Gyr. Hence, there is no remaining trace of a convective core visible in the seismic data at the current age of the star and the analysis is complete without explicit consideration of an overshoot in the modelling.

\section{Inversions for the Stellar Structure}
\label{sec_inversions}
In this section, we perform inversions on the CLES models computed in the previous part to improve the robustness of our analysis. Inversions are defined in such a way that they do not rely strongly on the choice of physical ingredients in the model and can thus extract additional informations from the asteroseismic data in a quasi model-independent way. Several prescriptions for the surface effects will be tested, by correcting the individual frequencies before the inversion or implementing them in the inversion. These inversions provide estimate of so-called indicators. There are several of them available in the literature \citep{2012A&A...539A..63R,2015A&A...574A..42B,2015A&A...583A..62B,2018A&A...609A..95B}, but we will limit ourselves to the mean density inversions. The $\tau$ indicator is indeed too sensitive to surface effects and the error on the correction predicted by the inversion is too high to extract meaningful informations with this quantity. In addition, some of the indicators that require numerous quadrupole modes \citep[e.g. $\rm S_{core}$ in][]{2018A&A...609A..95B} are not compatible with the limited number of observed quadrupole modes for Kepler-93.

\subsection{Theoretical Considerations}
\label{sec_theoretical_considerations}
The structure inversion equation is based on the analysis of the perturbation of the equation of motion describing the evolution of the displacement vector and considering only the linear terms. This approach is motivated by the work of \citet{1967MNRAS.136..293L} and their predecessors \citep[see e.g.][]{1964ApJ...139..664C,1964ApJ...140.1517C,1964ApJ...140.1045C}, who showed that the equation of motion fulfils a variational principle. In our case, the frequency perturbation is directly related to the structural perturbation and can be rewritten in the usual form~\citep{1990MNRAS.244..542D}:
\begin{equation}
\frac{\delta\nu^{n,l}}{\nu^{n,l}} = \int_{0}^{R} K_{a,b}^{n,l}\frac{\delta a}{a}dr + \int_{0}^{R} K_{b,a}^{n,l}\frac{\delta b}{b}dr + \mathcal{O}(\delta^2),
\label{eq_inversionGT}
\end{equation}
with $\nu$ the oscillation frequency, $a$ and $b$ two structural variables, $K_{a,b}^{n,l}$ and $K_{b,a}^{n,l}$ the structural kernels and using the definition:
\begin{equation}
\frac{\delta x}{x} = \frac{x_{\mathrm{obs}}-x_{\mathrm{ref}}}{x_{\mathrm{ref}}}.
\end{equation}
The index $ref$ stands for reference and $obs$ for observed. Then, the idea will be to invert equation \eqref{eq_inversionGT} and compute the indicator $t$ given the observed frequency differences.

The treatment of surface regions is quite approximative in equation \eqref{eq_inversionGT}, requiring the surface effects to be modelled by an additional empirical term, denoted $\mathcal{F}_{\mathrm{Surf}}$. In the literature \citep{2014A&A...568A.123B,2015A&A...583A.112S}, this term is considered to be slowly varying with the frequency and is determined in an empirical way. \citet{2014A&A...568A.123B} proposed a correction of the form:
\begin{align}
\delta\nu &= \left(a_{-1}\left(\frac{\nu}{\nu_{\mathrm{ac}}}\right)^{-1} + a_{3}\left(\frac{\nu}{\nu_{\mathrm{ac}}}\right)^{3}\right)/\mathcal{I},
\label{eq_BG2014_1}
\end{align}
where $\delta\nu$ is the estimated correction\footnote{$\delta\nu = \nu_{\mathrm{WSC}}-\nu_{\mathrm{NSC}}$, where WSC stands for with surface effects correction and NSC for no surface effects correction. By convention in this section, the frequency without index is the frequency without surface effects correction.}, $\mathcal{I}$ is the mode inertia and $a_{-1}$ \& $a_{3}$ are two coefficients to be added in the optimization procedure. The acoustic cut-off $\nu_{\mathrm{ac}}$ is computed using the scaling relation:
\begin{align}
\frac{\nu_{\mathrm{ac}}}{\nu_{\mathrm{ac},\odot}} = \frac{g}{g_\odot}\left(\frac{T_{\mathrm{eff}}}{T_{\mathrm{eff},\odot}}\right)^{-\frac{1}{2}},
\label{eq_BG2014_2}
\end{align}
with $g_\odot=27400$ cm/s$^2$, $T_{\mathrm{eff},\odot}=5772$ K and $\nu_{\mathrm{ac},\odot}=5000$ $\mu$Hz. On the other hand, the approach of \citet{2015A&A...583A.112S} is based on the stellar properties and does not require new optimization coefficients\footnote{The coefficients $\alpha$ and $\beta$ could be treated as free variables in the optimization process. However, since this correction is empirical, is it not worth to do so.}:
\begin{align}
\frac{\delta\nu}{\nu_{\mathrm{max}}} &= \alpha\left(1-\frac{1}{1+\left(\frac{\nu_{\mathrm{WSC}}}{\nu_{\mathrm{max}}}\right)^\beta}\right),
\end{align}
where $\alpha$ and $\beta$ can be determined from the surface gravity and effective temperature:
\begin{align}
\log|\alpha| &=\ \ \,  7.69\log T_{\mathrm{eff}} - 0.629\log g - 28.5 \\
\log\beta &= -3.86\log T_{\mathrm{eff}} + 0.235\log g + 14.2.
\end{align}

The frequency of maximal power $\nu_{\mathrm{max}}=3366$ $\mu$Hz is taken from \citet{2016MNRAS.456.2183D}.

In order to perform the inversion, we chose the Substractive Optimally Localized Averages (SOLA) approach \citep{1994A&A...281..231P}, which is an adaptation of the OLA approach of \citet{1968GeoJ...16..169B,1970RSPTA.266..123B}. Methods that takes into account the non-linearities can also be found in the literature \citep{2002ESASP.485...75R,2002ESASP.485..337R,2002ESASP.485..341R,2015A&A...582A..25A} and could be a good complement to confirm the results of this work. The SOLA method consists in the minimization of the following cost function for a given indicator $t$:
\begin{align}
\mathcal{J}_t(c_i) &= \int_0^1 \big(\mathcal{K}_{\mathrm{avg}} - \mathcal{T}_t\big)^2 dx 
                                        + \beta\int_0^1 \mathcal{K}_{\mathrm{cross}}^2dx &+ \lambda\left[2-\sum_i c_i\right] \nonumber\\
                                        &\quad+\tan\theta \frac{\sum_i (c_i\sigma_i)^2}{\langle\sigma^2\rangle}
                                        + \mathcal{F}_{\mathrm{Surf}}(\nu),
\end{align}
where the averaging and cross-term kernels are related to the structural kernels:
\begin{align}
\mathcal{K}_{\mathrm{avg}} &= \sum_i c_i K_{a,b}^{i} \\
\mathcal{K}_{\mathrm{cross}} &= \sum_i c_i K_{b,a}^{i}.
\end{align}
The variables $\beta$ and $\theta$ are trade-off parameters to adjust the balance between the amplitudes of the different terms during the fitting. The idea is to reduce the contribution of the cross-term and of the observational errors on the individual frequencies $\sigma_i$, while providing a good fit of the target function $\mathcal{T}_t$. A good fit of the target function by the averaging kernel ensures an accurate inversion result. Also, one defines $\langle\sigma^2\rangle = \sum_i^N\sigma_i^2$ with $N$ the number of observed frequencies. The inversion coefficients are denoted with $c_i$ and $\lambda$ is a Lagrange multiplier. The surface term $\mathcal{F}_{\mathrm{Surf}}(\nu)$ has to be treated with caution. Indeed, it allows to take into account the surface effects, however at the expense of the fit of the target function.

\subsection{Mean Density Inversions}
\label{sec_mean_density_inversions}
Mean density inversions were formalized  in \citet{2012A&A...539A..63R}. They are a well-tested method \citep{2012A&A...539A..63R,2015A&A...574A..42B}, which has been applied to various cases \citep{2016A&A...585A.109B,2016A&A...596A..73B,2019A&A...630A.126B,2021A&A...646A...7S} and that can efficiently extract key informations from asteroseismic observations in a quasi model-independent way. In addition, they do not necessarily require quadrupole modes which are more difficult to observe.

The form of the target function is motivated by considering the mass difference between the star and the reference model and deducing from it the relative difference in mean density:
\begin{align}
\frac{\delta\bar{\rho}}{\bar{\rho}} = \int_0^1 4\pi x^2 \frac{\rho}{\rho_R}\frac{\delta\rho}{\rho} dx,
\end{align}
with $\rho_R = \frac{M}{R^3}$ used for the non-dimensionalisation. $M$ is the stellar mass, $R$ its radius and $x=\frac{r}{R}$ the normalized radius. It follows that the target function of the mean density inversion is given by:
\begin{align}
\mathcal{T}_{\bar{\rho}} = 4\pi x^2 \frac{\rho}{\rho_R}.
\label{eq_target_function_rho}
\end{align}

The frequencies of the reference models for the inversion are computed without surface effects. We add them a posteriori, either by correcting the frequencies of the reference models and then performing the inversion or by implementing them in the SOLA cost function. We use four prescriptions to test their impact. The first prescription, labelled as \textit{no surface corr.}, is to carry out the inversion without surface effects. It is an extreme case that gives a lower bound for the mean density. The second, labelled as \textit{AIMS+BG 2014}, is to use the \citet{2014A&A...568A.123B} coefficients obtained with AIMS and the fitting of the individual frequencies. The third prescription, labelled as \textit{Sonoi 2015}, is to correct the modelled frequencies with the \citet{2015A&A...583A.112S} procedure. For these three prescriptions, inversions are carried without the term for the surface correction in the SOLA cost function. The last prescription, labelled as \textit{SOLA+BG 2014}, was to implement the \citet{2014A&A...568A.123B} correction in the $\mathcal{F}_{\mathrm{Surf}}$ term in the SOLA cost function. The downside of this approach is that it adds two free variables in the minimization. Consequently, the target function was less well reproduced by the averaging kernel and the uncertainty on the inverted mean density is significantly higher.

\begin{figure}
\centering
\includegraphics[scale=0.35]{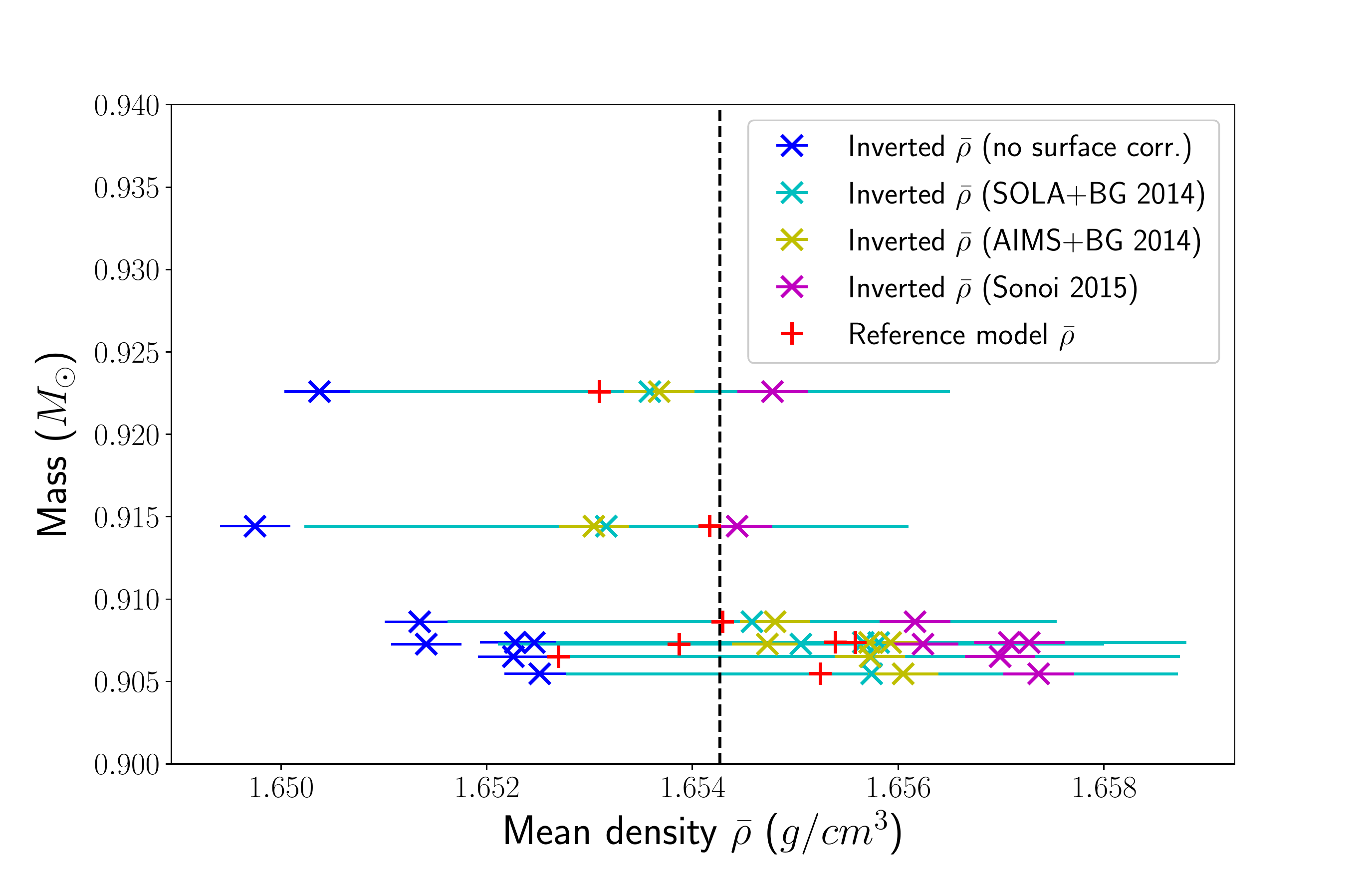}
\caption{Mean density inversions results. The vertical black dashed line is the weighted mean. The four prescriptions for the surface effects are described in details in the paragraph right after Eq. \eqref{eq_target_function_rho}. \textit{no surface corr.} corresponds to the first prescription, \textit{AIMS+BG 2014} to the second, \textit{Sonoi 2015} to the third and \textit{SOLA+BG 2014} to the last prescription.}
\label{fig_results_inversion}
\end{figure}

\begin{figure}
\centering
\includegraphics[scale=0.45]{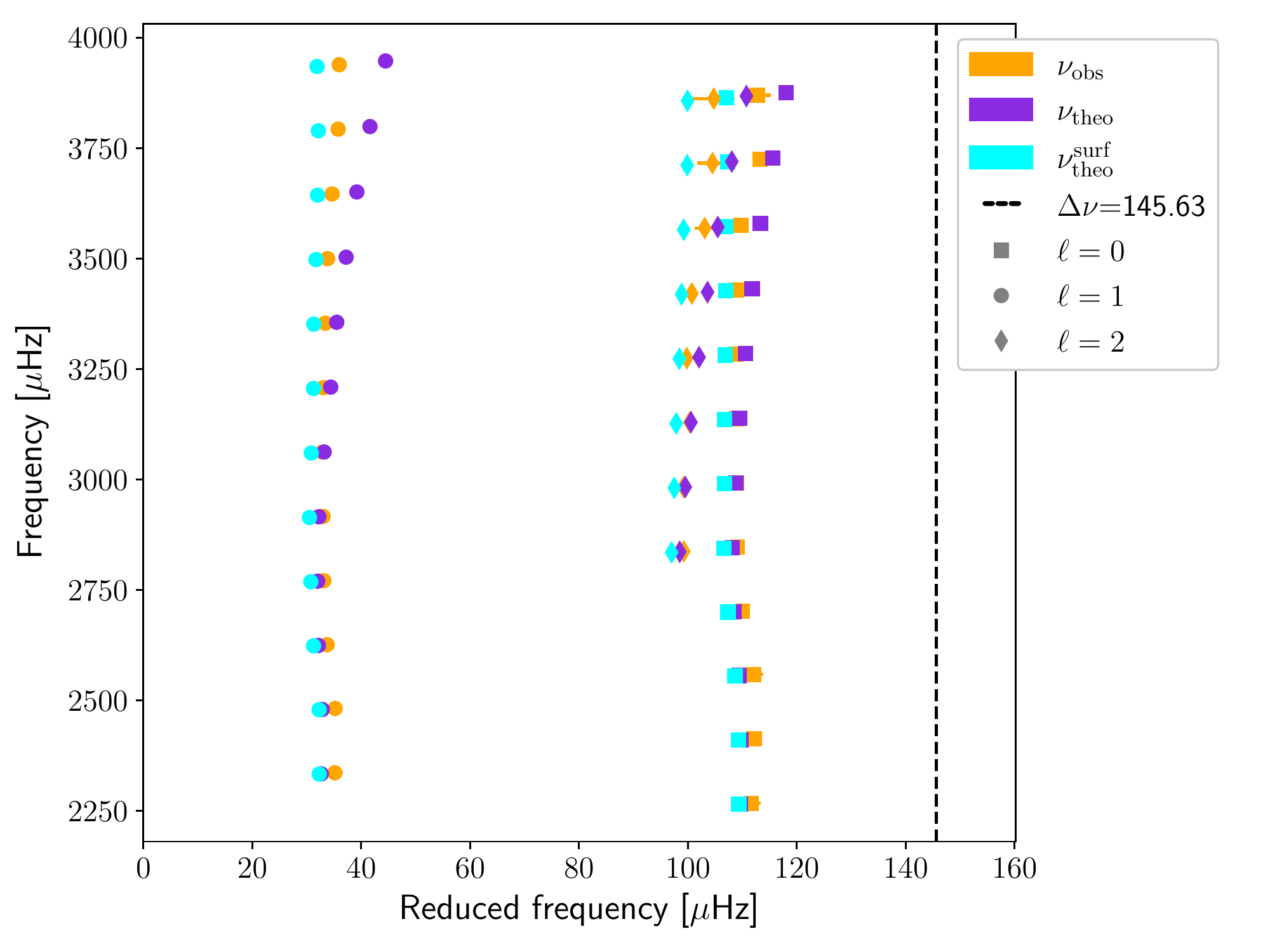}
\caption{Echelle diagram of the result of AIMS with the fit of the ratios. The inverted mean density was part of the constraints. The comparison between the results with (cyan) and without (purple) correction for the surface effect is shown. The \citet{2014A&A...568A.123B} correction was chosen for this plot, but the effect is similar with another prescription for the surface effects.}
\label{fig_comparison_surface_effects_with_without}
\end{figure}

The results of the inversions for the different models are displayed in Fig. \ref{fig_results_inversion}. This plot shows the necessity to take into account the surface effects. In fact, all the uncorrected models are biased and lie on the left of the plot (dark blue points). The results with the implementation of the surface effects are also not totally bias-free. Indeed, as illustrated in Fig. \ref{fig_comparison_surface_effects_with_without}, the model without surface effects tends to overestimate the observed high frequencies, where the contribution of these surface effects is stronger, while the frequencies estimated taking into account the surface effects tend to slightly underestimate the frequencies.

The revised stellar parameters are listed in Table \ref{tab_revised_stellar_parameters}. The mass, radius and age come from the fit of the ratios with AIMS. The uncertainties are the addition of the statistical uncertainties provided by AIMS and the systematic errors due to the choice of physical ingredients and of observables\footnote{The observables refer to the seismic constraints used in the fit: individual frequencies or frequency ratios.} estimated with the standard deviation of the models listed in Table \ref{tab_parameters_levenberg}. We did not include the models with overshoot in the standard deviation as they predicted frequency ratios that were in worse agreement with the observations. The standard deviation is therefore based on the models 1-2 \& 7-12. In order to provide a consistent estimate of the stellar mean density, we computed the weighted mean\footnote{We chose our weights to be $w_i=1/\sigma_i^2$.} of the inversion results with and without surface effects, since the true value is expected to lay in-between. Note that the weighted mean and the median are almost equal. The error is chosen in such a way that the majority of the models are within the error interval. This conservative estimation of the uncertainty still leads to a great precision on the mean density, of about 0.2\%.

A consistency test is then conducted. A mass range is deduced from the mean density and using the Stefan-Boltzmann law to get the radius from the spectroscopic constraints:
\begin{align}
M_{\mathrm{SB}} &= \frac{4\pi}{3}\bar{\rho}\left(\frac{L}{4\pi\sigma_{\mathrm{SB}}T_{eff}^4}\right)^{3/2} \\
\sigma_{M_{\mathrm{SB}}} &= M_{\mathrm{SB}}\sqrt{\left(\frac{\sigma_{\bar{\rho}}}{\bar{\rho}}\right)^2 + \left(\frac{3}{2}\frac{\sigma_L}{L}\right)^2 + \left(6\cdot\frac{\sigma_{T_{eff}}}{T_{eff}}\right)^2}.
\end{align}  
This mass range is based on the spectroscopic constraints and the mean density obtained with the inversions. It is therefore quasi independent of the forward modelling. One gets $M_{\mathrm{SB}}=0.925\pm 0.110~M_\odot$. All the mass estimates of our work and of the literature lie within this mass range, what is consistent with our expectation. The precision on $M_{\mathrm{SB}}$ is limited by the error on the effective temperature. Indeed, the best estimation of the effective temperature found in the literature was computed within a survey and could thus be improved with a detailed spectroscopic analysis of Kepler-93. This point is crucial for improvements, since the estimation of the effective temperature also affects the estimated stellar luminosity. For example, if the error on the effective temperature was 50 K instead of 100 K, one could reduce the error on $M_{\mathrm{SB}}$ by about 35\%.

The stellar parameters are consistent with the literature values. Notice that \citet{2015MNRAS.452.2127S} also investigated the impact of input physics, but they focussed on the abundances and the mixing-length parameter. In our work, we tested many more physical ingredients (abundances, opacities, turbulent diffusion, overshoot, ...). In addition, \citet{2015MNRAS.452.2127S} considered a sample of 33 targets and estimated the systematics based on median values, while we conducted a detailed analysis on a single star. Since their sample contains F-type stars and stars massive enough to have a convective core, their uncertainties are therefore likely overestimated in the case of solar-type stars, such as Kepler-93. This especially explains why they quote a precision on the age that is significantly lower than ours. They also did not used mean density inversions, thus limiting the precision on the mean density in their forward modelling. Finally, we remark that our detailed analysis of Kepler-93 provided stellar parameters that are well within the precision requirements for the PLATO mission.

\begin{table}
\centering
\caption{Revised stellar parameters of Kepler-93. The quoted errors include the systematics due to the choice of the physical ingredients for the mass, age and radius and the impact of surface effects and model-dependency in the inversion for mean density.}
\resizebox{\linewidth}{!}{
\begin{tabular}{cccc}
\hline 
\textbf{Mean density} (g/cm$^3$) & \textbf{Mass} (M$_\odot$) & \textbf{Radius} (R$_\odot$) & \textbf{Age} (Gyr) \\ 
\hline \hline 
$1.654\pm 0.004$ & $0.907\pm 0.023$ & $0.918\pm 0.008$ & $6.78\pm 0.32$ \\ 
\hline 
\end{tabular} 
}
\label{tab_revised_stellar_parameters}
\end{table}

\subsection{Revision of the Planetary Parameters}
\label{sec_planetary_parameters_revision}
Based on the improvement on the precision on the stellar parameters, we provide in Table \ref{tab_revision_kepler93b} a revision of the planetary parameters of Kepler-93b.

The planet induces a radial velocity signal on the host star, what allows to find the planetary mass \citep{1999ApJ...526..890C}:
\begin{equation}
\label{eq_mass_planet}
K = 	\left(\frac{2\pi G}{P}\right)^{1/3}\frac{M_p}{(M_*+M_p)^{2/3}}\frac{\sin i}{\sqrt{1-e^2}},
\end{equation}
where $K$ is the RV amplitude, $G$ the Newtonian gravitational constant, $P$ the orbital period of the planet, $M_*$ the host star mass, $M_p$ the planet mass, $i$ the inclination and $e$ the orbital eccentricity. As in \citet{Dressing2015}, we assumed that $e=0$, since they found that a non-zero eccentricity did not provide a better fitting in their analysis. Although this equation possesses an exact solution using an appropriate substitution and Cardano's formula, it is numerically more stable to use a Monte-Carlo procedure to estimate the error on the planetary mass. At first, we investigated the impact of the precision on the stellar mass by letting the other parameters fixed. We found that the factor two of improvement leads to a factor two on the precision of the planet mass. However, the RV amplitude is not error-free and turns out to completely dominate the uncertainty on the planetary mass. The main contributor on the error on the planet mean density is the planet mass. Consequently, the precision on this quantity remains unchanged. This issue will be less important for PLATO or TESS targets observed in the Southern hemisphere for which the RV follow-up will be more thorough.

The gain in precision on the stellar radius is more significant, especially as it improves the uncertainty on the planetary radius and orbital distance by about 25\% and 27\% with respect to the values of \citet{Dressing2015}.
\begin{table}[h]
\centering
\caption{Revised planetary parameters of Kepler-93b. The uncertainties of the values revised in this work include the systematics found for the stellar parameters.}
\resizebox{\linewidth}{!}{
\begin{tabular}{lcc}
\hline
Parameter & Value with $1\sigma$ error & Reference \\ 
\hline \hline
\textit{Transit and orbital parameters} & & \\ 
Orbital period $P$ (days) & $4.72673978 \pm 9.7 \times 10^{-7}$ & 1 \\ 
$R_p/R_*$ & $0.014751 \pm 0.000059$ & 1 \\ 
$a/R_*$ & $12.496 \pm 0.015$ & 1 \\  
Inclination $i$ (deg) & $89.183 \pm 0.044$ & 1 \\ 
Orbital eccentricity $e$ & 0 (fixed) & 2 \\ 
RV semi-amplitude $K$ (m/s) & $1.63 \pm 0.27$ & 2 \\ 
\hline \hline
\textit{Planetary parameters} &  &  \\  
$R_p (R_\oplus)$ & $1.478 \pm 0.014$ & 3 \\ 
$M_p (M_\oplus)$ & $4.01\pm 0.67$ & 3 \\  
$\rho_p$ (g/cm$^3$) & $6.84 \pm 1.16$ & 3 \\  
$\log g_p$ (cgs) & $3.256\pm 0.074$ & 3 \\ 
$a$ (AU) & $0.0533\pm 0.0005$ & 3 \\   
\hline
\multicolumn{3}{l}{\textbf{References.} (1) \citet{2014ApJ...790...12B}; (2) \citet{Dressing2015}; (3) this work.} \\
\end{tabular}
}
\label{tab_revision_kepler93b}
\end{table}

\section{Orbital Evolution of Kepler-93b}
\label{sec_Kepler93b}
Thanks to the detailed asteroseismic characterisation of the system presented in the previous sections, which provided very precise stellar parameters and revised planetary parameters, we may proceed with studying the orbital evolution of Kepler-93b. In this section, in particular we investigate whether the gravitational tidal interaction between the planet and the host star played a significant role in the past history of the system, and consequently we estimate the X-ray and EUV fluxes received by the planet during its potential orbital motion.

In this context, the availability of precise parameters (such as the age) is crucial in order to discard past evolutionary scenarios which would prevent us from reproducing the current status of the system within the uncertainties of the fundamental quantities. Therefore we started our study by computing a model of Kepler-93 by means of the CLES stellar evolution code, using the stellar parameters derived above as input and constraints for the model.

In order to follow the evolution of the planetary system, we couple the stellar model to our orbital evolution code \citep{Privitera2016A, Privitera2016B,Rao2018,2021A&A...650A.108P}, taking into account the exchange of angular momentum between the star and the orbit, once the protoplanetary disk has dissipated. Thanks to this approach, we can test whether dynamical and/or equilibrium tides have significantly impacted the orbit of the planet. Dynamical tides are mainly efficient during the PMS phase when the host star rotates faster, while equilibrium tides take over at the later stages of the evolution. The physics included in the orbital evolution code is the one described in \citet{Rao2018} and \citet{2021A&A...650A.108P}. Estimations of the mass loss from planetary atmospheres are provided following the formalism of the Jeans or hydrodynamic escape regimes, depending on the properties of the planetary system considered.

The rotational history of Kepler-93 being unknown, we consider three different values for the initial surface rotation rate, representative of slow ($\rm \Omega_{in} = 3.2\times \Omega_{\odot}$), medium ($\rm \Omega_{in} = 5\times \Omega_{\odot}$) and fast rotators ($\rm \Omega_{in} = 18\times \Omega_{\odot}$) as deduced from surface rotation rates of solar-type stars observed in open clusters at different ages \citep{Eggenberger2019a}. A disk lifetime $\rm \tau_{dl} =6~Myr$ is adopted for medium and slow rotators, while $\rm \tau_{dl} =2~Myr$ is adopted for the fast rotator. During the disk-locking timescale, the surface rotation of the host-star is assumed to remain constant. After disk dispersal, the star is supposed to rotate as a solid body. This assumption receives some support from the the rotational profile deduced for the Sun from helioseismology \citep[see e.g.][]{Kosovichev1988,Brown1989,Elsworth1995,Kosovichev1997,Couvidat2003} that is flat in most of the envelope and the nearly-uniform internal rotation of solar-like main sequence stars derived from seismic analysis \citep[see e.g.][]{Lund2014,Benomar2015}.

For the magnetic braking of the stellar surface we refer to the formalism of \citet{Matt2015,Matt2019}, for which the torque is given by

\begin{equation}
\rm \dfrac{dJ}{dt} =
\begin{cases}
\rm -T_{\odot} \left(\dfrac{R_\star}{R_{\odot}} \right)^{3.1} \left( \dfrac{M_\star}{M_{\odot}} \right)^{0.5} \left(\dfrac{\tau_{cz}}{\tau_{cz \odot}} \right)^{p} \left(  \rm \dfrac{\Omega}{\Omega_{\odot}} \right)^{p+1} &, \rm \text{if} ~ \left( Ro > Ro_{\odot}/\chi \right),\\
\rm -T_{\odot} \left(\dfrac{R_\star}{R_{\odot}} \right)^{3.1} \left( \dfrac{M_\star}{M_{\odot}} \right)^{0.5} \chi^{p} \left( \dfrac{\Omega}{\Omega_{\odot}} \right) & , \rm \text{if} ~ \left( Ro \leq Ro_{\odot}/\chi \right) ,
\end{cases}
\end{equation}

where $\rm R_\star$ and $\rm M_\star$ are the radius and the mass of the stellar model, $\rm R_{\odot}$ and $\rm M_{\odot}$ are the radius and the mass of the Sun. The convective turnover timescale is indicated as $\rm \tau_{cz}$ and $\rm Ro$ is the Rossby number, defined as the ratio between the stellar rotational period and the convective turnover timescale ($\rm Ro = P_{\star}/\tau_{cz}$). The term $\rm \chi \equiv Ro_{\odot} / Ro_{sat}$ indicates the critical rotation rate for stars with given $\rm \tau_{cz}/\tau_{cz_{\odot}}$, defining the transition from saturated to unsaturated regime. Here we take $\rm \chi = 10$ as in \citet{Matt2015} and \citet{Eggenberger2019a}. The exponent $\rm p$ is taken equal to $\rm 2.3$ and the constant $\rm T_{\odot}$ is calibrated in order to reproduce the solar surface rotation rate \citep{Eggenberger2019a}. \\

\begin{figure}
\centering
\includegraphics[width=0.5\textwidth]{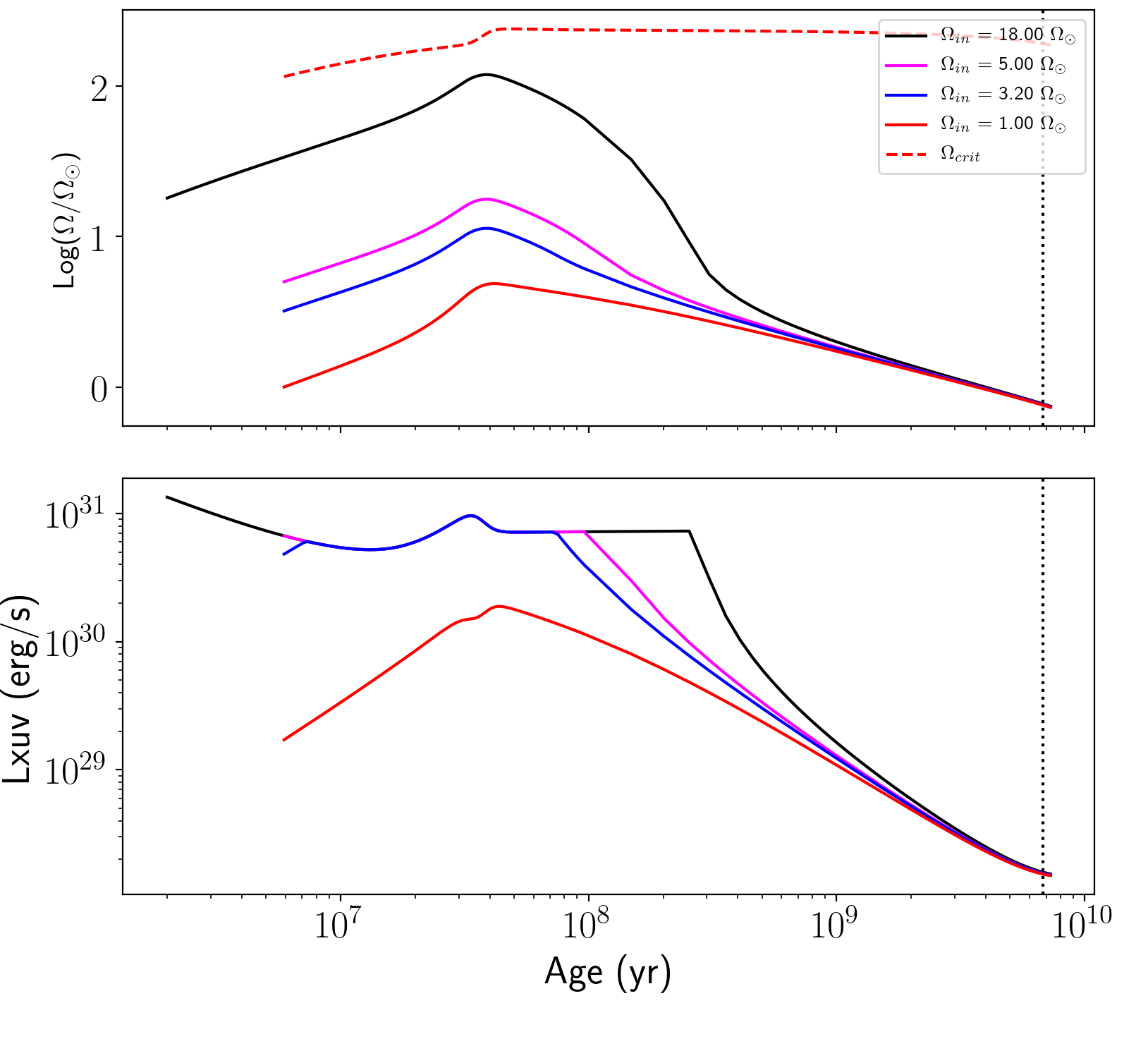}
\caption{\footnotesize{\emph{Top panel:} evolution of the surface rotation rate of Kepler-93, in case of fast rotator ($\rm \Omega_{ini} = 18 \times \Omega_{\odot}$, black solid line), medium rotator ($\rm \Omega_{ini} = 5 \times \Omega_{\odot}$, magenta solid line), slow rotator ($\rm \Omega_{ini} = 3.2 \times \Omega_{\odot}$, blue solid line) and super-slow rotator ($\rm \Omega_{ini} = \Omega_{\odot}$, red solid line). The red dashed line shows the critical velocity limit. The black-dotted vertical line indicates the age of the system. \emph{Bottom panel:} XUV luminosity evolutionary tracks of Kepler-93, relatively to the different rotational histories considered.}}
\label{Fig:rotational_history}
\end{figure}

In the top panel of Fig.~\ref{Fig:rotational_history}, we present the evolution of the surface rotation rates computed for Kepler-93, starting from the dispersal of the protoplanetary disk until the age of the system. We also show the evolution of the surface rotation rate of a super-slow rotator ($\rm \Omega_{in} = \Omega_{\odot}$). While such a rotator is less representative of the distribution of surface rotation rates observed for stars in open clusters at different ages, we include it in our study being interested in considering its impact on the evaporation of the planetary atmosphere (see below). Unfortunately, there is no precise estimation of the surface rotation rate of Kepler-93 to compare our models with. \citet{Mazeh2015} provide a value of the rotational period of the star, but they flag the detection as inconsistent in different quarters and not robust. \citet{Suto2019} also estimated a rotation period for Kepler-93, but they considered this star has having no clear signal in its periodogram and flagged the period as unreliable. At the age of the system, our models predict a value of the rotational period $\rm P_{rot} = 33.7~d$.

In the bottom panel of Fig.~\ref{Fig:rotational_history}, we show the XUV-luminosity evolutionary tracks computed for fast (solid black line), medium (solid magenta line), slow (solid blue) and super-slow rotators (solid red line). Following the work by \citet{Tu2015}, we compute the emission of the X-ray luminosity recalibrating the prescription of \citet{Wright2011}. Their prescription gives the ratio of the X-ray luminosity with respect to the bolometric luminosity of the star ($\rm Rx = Lx/L_{\star}$) in saturated and unsaturated regimes, depending on the value of the Rossby number, defined as the ratio between the stellar rotational period and the convective turnover timescale. For the computation of the EUV luminosity, we use the prescription of \citet{SanzForcada2011}. Among the rotators considered, only the super-slow one never enters the saturation regime, providing a XUV luminosity emission that occurs at globally lower magnitudes.

We start our computations by studying the impact of dynamical tides on the total change of the orbital distance. We first choose the current value of the orbital distance as initial input for our computations ($\rm a_{in} = 0.0533~AU$, Tab.~\ref{tab_revision_kepler93b}), in order to test whether we reproduce the position of the planet at the age of the system. For the planetary mass, we take the maximal value allowed for Kepler-93b within the error bar ($\rm M_{in} = 4.68~M_{\oplus}$, Tab.~\ref{tab_revision_kepler93b}) and consider it to remain constant during its evolution. In order to maximize the impact of dynamical tides, we consider Kepler-93 having started the evolution as a fast rotator, namely with an initial surface rotation rate $\rm \Omega_{in} = 18 \times \Omega_{\odot}$. As a result of this simulation, we find that tides do not play a significant role in shaping the architecture of the system, leaving the orbit of the planet unperturbed. 

\begin{figure}
\centering
\subfigure{\includegraphics[width=0.48\textwidth]{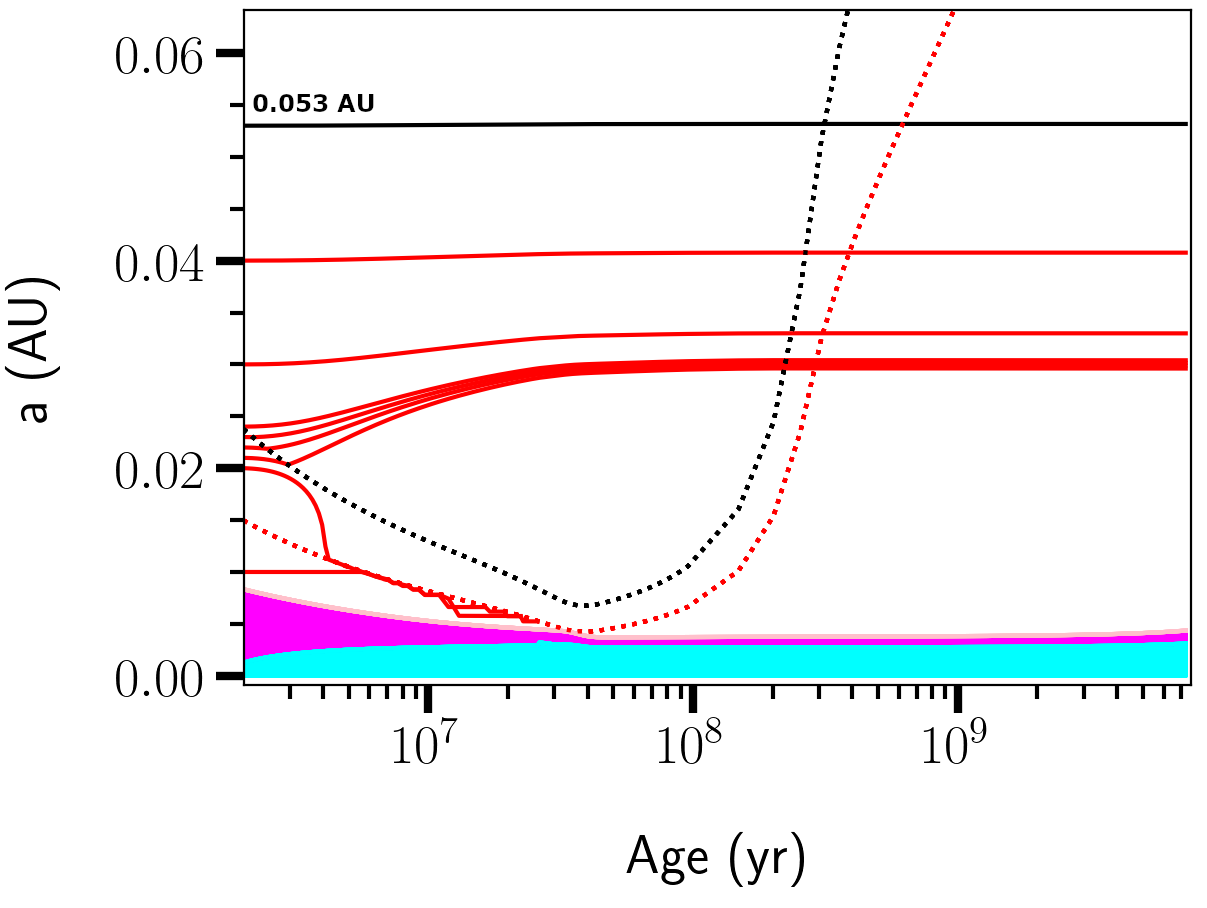}}
\subfigure{ \includegraphics[width=0.48\textwidth]{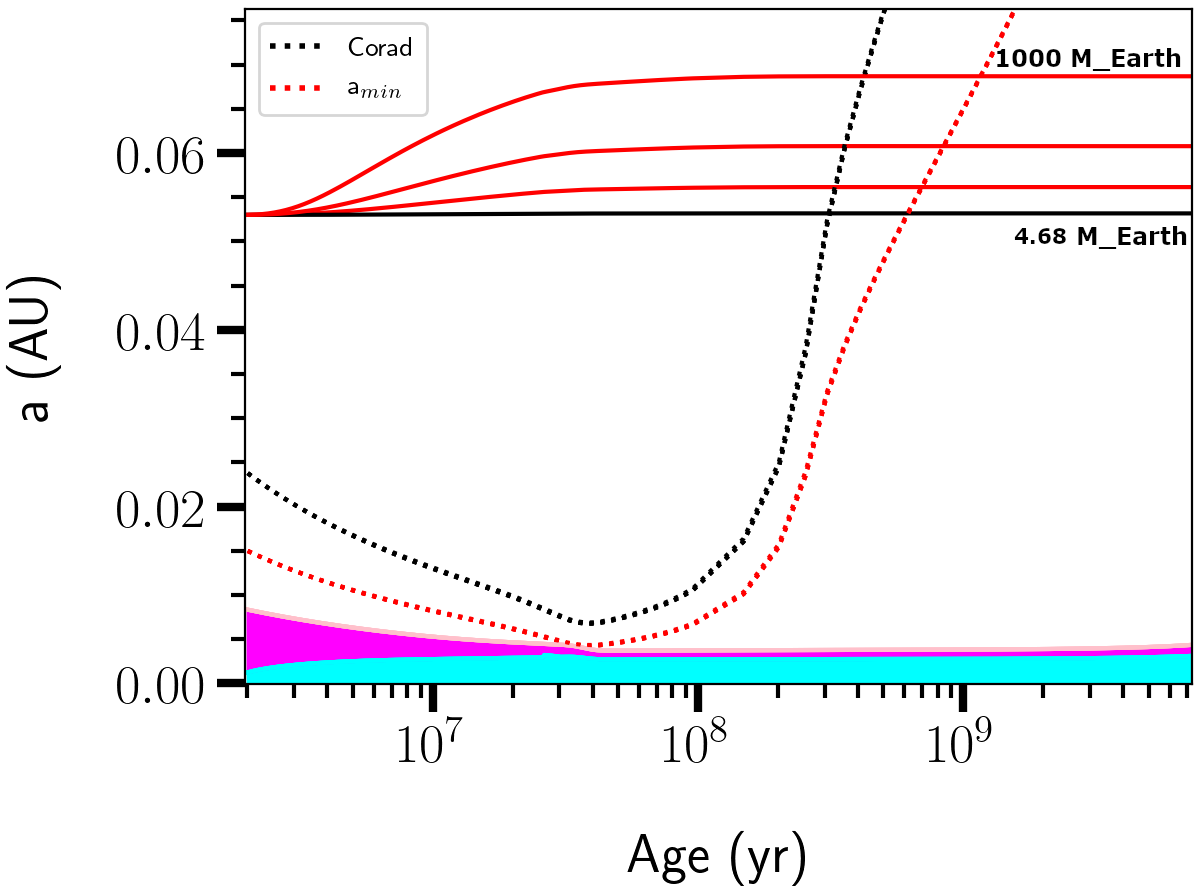}}
\caption{\footnotesize{\emph{Top panel:}  orbital evolution of a planet having a mass fixed at the maximal value allowed for Kepler-93b ($\rm M_{in} = 4.68~M_{Earth}$), for different values of the initial orbital distance: $\rm a_{in} = 0.01, 0.02, 0.021, 0.022, 0.023, 0.024, 0.03, 0.04$ and $\rm 0.053$ AU. For initial orbital distances lower than $\rm \sim 0.023~AU$, the planet would be rapidly engulfed by the host star. The initial surface rotation rate considered in both panels is $\rm \Omega_{in} = 18 \times \Omega_{\odot}$. \emph{Bottom panel:} Orbital evolution of a planet located at the same orbital distance of Kepler-93b (initial value fixed at $\rm a_{in} = 0.053~AU$), by considering different initial masses, the maximal value allowed for Kepler-93b ($\rm M_{in} = 4.68~M_{\oplus}$, solid black line) and $\rm M_{in} = 100, 317, 1000~M_{\oplus}$ (red solid lines). The magenta area represents the extension of the stellar convective envelope, while the cyan area represents the radiative core. The dotted black lines indicate the evolution of the corotation radii and the dotted red lines show the evolution of the minimum orbital distance for dynamical tides to be active.}}
\label{Fig:test orbits}
\end{figure}

We note that the relatively low mass ratio between the planet and the star, together with the large value of the initial orbital distance, play a significant role in determining the efficiency of tides. We investigate these points in more details by doing some additional computations in order to verify how sensitive this result depends on the values of the initial planetary mass and orbital distance.
Firstly, we study the impact of changing the value of the initial orbital distance. We compute orbital evolutions for a planet with a mass fixed to the maximal value allowed for Kepler-93b ($\rm M_{in} = 4.68~M_{\oplus}$) and semi-major axis lower than $\rm 0.0533~AU$. In the top panel of Fig.~\ref{Fig:test orbits}, the orbital evolutions corresponding to different values of the semi-major axis are presented. We notice that for values of $\rm a_{in}$ ranging between roughly $\rm 0.022 - 0.04~AU$ the orbits expand, while for $\rm a_{in} \leq 0.02~AU$ the planet is rapidly engulfed by the host star ($\rm t_{eng} \sim 3 \times 10^7~ Myr$). Notably, none of the values of the semi-major axis considered would reproduce the orbital distance of Kepler-93b at the age of the system. On the base of this result, we would consider the planet having evolved at a fixed orbital distance ($\rm \sim 0.0533~AU$), after the protoplanetary disk dispersal. 
In a similar way, given the uncertainty on the measurement of the mass of Kepler-93b, we compute the orbital evolution of a planet having the same initial orbital distance of Kepler-93b, but higher initial mass. As it is shown in the bottom panel of Fig.~\ref{Fig:test orbits}, we find a slight increase of the orbit only for planets having an initial mass significantly larger than one estimated for Kepler-93b within the errorbar, namely only for  $\rm M_{pl} \gtrsim  100~M_{\oplus}$. This result indicates that if Kepler-93b formed with a mass similar to its present value, tides did not play a crucial role in shaping the orbit of this planet.

We may ask whether Kepler-93b could have formed with an initial mass as high as $\rm 100~M_{\oplus}$. This hypothesis would require that the planet experienced a substantial mass loss during its evolution in order to end up with only $\rm \sim 4~M_{\oplus}$ at the age of the system. Such a planet might also have started the evolution with an orbital distance lower than $\rm 0.0533$ AU, given that dynamical tides have been proven to expand the orbits of planets as massive as $\rm 100~M_{\oplus}$ on timescales of the order of $\rm 30$ Myr. We therefore recomputed the evolution of the planetary orbit, this time considering both the impact of the planetary mass-loss through photoevaporation, and the one of tides. We explored a range of initial orbital distances between $\rm 0.02$ and $\rm 0.0533$ AU, for initial planetary masses $\rm M_{in} = 100, 317, 1000~M_{\oplus}$. Increasing the initial mass of the planet results in the more efficient impact of dynamical tides. Therefore, the maximum expansion of the orbit is obtained for the planet having $\rm M_{in} = 1000~M_{\oplus}$. We found that for some of the $\rm a_{in}$ values considered we are able to reproduce the current position of the planet. However, the mass-loss obtained through photoevaporation is not strong enough to efficiently remove a significant percentage of the planetary mass. We find that for the initial masses considered, the escape process occurs in Jeans escape regime conditions. This result was found for a star having an initial surface rotation rate $\rm \Omega_{in} = 18 \times \Omega_{\odot}$. Lower values of $\rm \Omega_{in}$ would lead to similar outcomes, with a less efficient impact of tides. Therefore, according to our results it seems unlikely that a planet like Kepler-93b formed with a mass large enough ($\rm M_{in} \gtrsim 100~M_{\oplus}$) to see its orbit affected by stellar tides, in the hypothesis that photoevaporation is the only process inducing mass-loss from the planet.\\

The relatively high density estimated for Kepler-93b ($\rho = 6.84 \pm 1.15$ g/cm$^3$) would characterise this planet as a rocky world \citep{2014ApJ...790...12B, Dressing2015, Dorn2017}. If there was an atmosphere surrounding the planet after the protoplanetary disk dispersal, it may have been lost at the early stages of the evolution through different processes, among which core-powered mass loss and/or evaporation due to high-energy stellar photons \citep{Owen2017,Ginzburg2018}. 
Using photoevaporation, we aim at estimating the maximum initial mass of the planet ($\rm M_{max}$) above which it would be able to retain a H/He rich atmosphere after $\sim 6.78$ Gyr. We estimate the value of $\rm M_{max}$ integrating backward in time the mass-loss rates obtained by using both the energy limited formula \citep{Erkaev2007, Lecavelier2007}, for which we assume a heating efficiency $\rm \eta = 0.15$, and the hydro-based approximation of \citet{Kubyshkina2018}. We computed the mass-loss rates for the fast and super-slow rotator. When using the energy-limited formalism, we obtain that $\rm M_{max}$ is $\rm 9.4$ and $\rm 5.4~M_{\oplus}$, for the fast and super-slow rotator respectively. In the case of the hydro-based approximation instead, we get a significantly larger value of $\rm  26.5~M_{\oplus}$ for the fast rotator and $\rm 8.3~M_{\oplus}$ for the slow one.

\section{Conclusions}
\label{sec_conclusion}
In this study, we carried out a detailed modelling of Kepler-93, a solar-like exoplanet host star observed continuously for almost the whole duration of the nominal \textit{Kepler} mission. We used a combination of seismic, spectroscopic and astrometric constraints to determine a best-fitting evolutionary model in Sect. \ref{sec_forward_modelling} using a wide range of physical ingredients for the model, thus providing a clearer view of the contributions coming from modelling uncertainties in comparison to those of observational data. This first step of seismic forward modelling is supplemented in Sect. \ref{sec_inversions} by a seismic inversion of the mean density, following \citet{2012A&A...539A..63R}, where we have tested the impact of both model-dependency and surface effects on the inversion results. In Sect. \ref{sec_Kepler93b}, we used this model to study the orbital evolution under the effects of equilibrium and dynamical tides and the evaporation of the planetary atmosphere from the effects of the XUV flux. 

Our final reported stellar parameters are $\bar{\rho}_*=1.654\pm0.004$~g/cm$^{3}$, $M_*=0.907\pm0.023\ M_{\odot}$, $R_*=0.918\pm0.008\ R_{\odot}$ and $\rm{Age}=6.78\pm0.32~\rm{Gyr}$ and agree with previous determinations in the literature. The results of this detailed modelling have been used to revise the planetary parameters of Kepler-93b, $\bar{\rho}_p=6.84\pm 1.16$ g/cm$^{3}$, $M_p=4.01\pm0.67\ M_{\oplus}$, $R_p=1.478\pm0.014\ R_{\oplus}$, $\log g_p = 3.256\pm 0.074$ and $a=0.0533\pm 0.0005$ AU. We also see that at the level of precision of \textit{Kepler} data, the spread obtained from variation of microphysical ingredients such as abundances, opacities, formalism of microscopic diffusion, as well as the inclusion of macroscopic transport at the base of the envelope in the form of turbulent diffusion does not impact significantly the mass and radius estimates. This is different from what was obtained in the case of Kepler-444 \citep{2019A&A...630A.126B}, where the spread in mass and radius was significant at the level of precision of the Kepler data. However, we see that the age determination of Kepler-93 can be significantly affected by a revision of the radiative opacities. In our case, switching from the OPAL \citep{1996ApJ...464..943I} to the OPLIB \citep{2016ApJ...817..116C} opacities induces a change of $\approx 4\%$, which is not negligible at our reported precision. In the context of the solar modelling problem \citep[See e.g.][and references therein]{2008PhR...457..217B,2020arXiv200706488C,2019FrASS...6...42B}, this emphasizes that any revision of the opacities might impact the determined ages from stellar evolutionary models in the case of solar-like stars. Nevertheless, the overall spread we see in the fundamental stellar parameters is rather small, and within the requirements of the PLATO mission (2\% for the radius, 15\% for the mass and 10\% for the age), even when taking into account variations of physical ingredients within the current limitations of the stellar models.

While the effects of physical ingredients are in this specific case more restricted, the variations induced by choosing another set of seismic constraints are not. This can be seen from Tables \ref{tab_AIMS} and \ref{tab_parameters_levenberg} where the use of individual frequencies as direct seismic constraints leads to a larger mass by about $\approx 1.5\%$ and radius by about $\approx 0.5\%$ than the one determined from the fit of the frequency ratios. This is in line with what is shown by 	\citet{2020MNRAS.495.4965J} when testing various surface corrections in the seismic forward modelling using AIMS (although for red-giant branch stars). In that respect, the choice of the seismic constraints for the modelling has also an impact on the final set of parameters, at least at the precision aimed for in the context of the PLATO mission.

We also tested the possibility of the survival of a convective core coming from out-of-equilibrium $^3$He burning at the beginning of the evolution, as seen in HD203608 \citep{2010A&A...514A..31D}. While Kepler-93 has a similar mass, it has a much higher metallicity $\left[\rm{Fe/H} \right]=-0.18\pm0.1$ while HD203608 has $\left[\rm{Fe/H} \right]=-0.5 \pm 0.1$, thus the opacity in the deep layer could be different and thus impact the size and survival of the convective core. Using the $r_{01}$ frequency ratios in Fig. \ref{fig_ratios_comp}, we can see that sustaining the out-of-equilibrium burning for about 3 Gyr leads to a worse agreement with seismic data. This implies that a convective core could not have been long-lived in Kepler-93, unlike in Kepler-444 and HD203608.

We computed stellar models of Kepler-93 using the stellar parameters derived from the asteroseismic characterisation as inputs and constraints. The rotational history of Kepler-93 being unknown, we considered different values for the initial surface rotation rates, representative of slow, medium and fast rotators. We studied the orbital evolution of Kepler-93b coupling the stellar models to our orbital evolution code, investigating the combined impact of dynamical/equilibrium tides and atmospheric photoevaporation. Exploring a range of initial orbital distances and planetary masses, we found that it is unlikely that Kepler-93b could have formed with a mass large enough to see its orbit perturbed by stellar tides, if we assume that photoevaporation is the only process inducing mass-loss from the planet.

In the context of the preparation of the PLATO mission, our study proves that for the benchmark target of a solar-like star with a visible magnitude of 10, a precision of 10\% in age is achievable with a detailed seismic modelling procedure and the use of mean density inversions. We show that this threshold of 10\% is reached if a precision of around 3\% in mass and 1\% in radius is achieved. However, the variations of 4\% in age observed when using the OPLIB opacities still advocate for caution regarding the actual accuracy of the stellar models. In that respect, further tests for benchmark stars are required to test the impact of metallicity, especially for cases deviating more significantly from the solar metallicity. Our study also underlines the relevance of inversion techniques for the PLATO mission. Indeed, the inclusion of the mean density in the constraints can improve the precision on the stellar mass and radius. Inversions are feasible as long as individual acoustic oscillations are determined for a given target. This is expected to be the case for the majority of the PLATO sample of main-sequence solar-type stars ($\sim$ 10000 stars). For red giants and subgiants, mixed modes can exhibit intrinsic non-linearities but mean density inversions have been shown to be feasible for such targets using radial modes only \citep{2019MNRAS.482.2305B}. Hence, inversions will hopefully be of a great help for the majority of the PLATO sample.

\section*{Acknowledgements}
J.B and G.B. acknowledge fundings from the SNF AMBIZIONE grant No 185805 (Seismic inversions and modelling of transport processes in stars). C.P. acknowledges fundings from the Swiss National Science Foundation (project Interacting Stars, number 200020-172505). P.E. and S.J.A.J.S. have received funding from the European Research Council (ERC) under the European Union's Horizon 2020 research and innovation programme (grant agreement No 833925, project STAREX). A.M. acknowledges support from the ERC Consolidator Grant funding scheme (project ASTEROCHRONOMETRY, G.A. No 772293).

\bibliography{bibliography.bib}

\appendix

\section{Observational Data}
\label{sec_appendix_observational_data}
\begin{table}[h!]
\caption{Observational data of Kepler-93. The observed frequencies are from \citet{2016MNRAS.456.2183D}. We corrected the mode identification by one radial order, as explained in Sect. \ref{sec_forward_modelling}.}
\begin{tabular}{cccc}
\hline
$l$ & $n$ & Frequency & 68\% credible \\
 & & ($\mu$Hz) & ($\mu$Hz) \\
\hline\hline
0 & 14 & 2266.65 & 1.78 \\
0 & 15 & 2412.81 & 0.53 \\
0 & 16 & 2558.34 & 1.79 \\
0 & 17 & 2701.90 & 0.19 \\
0 & 18 & 2846.59 & 0.14 \\
0 & 19 & 2992.05 & 0.11 \\
0 & 20 & 3137.69 & 0.12 \\
0 & 21 & 3283.18 & 0.09 \\
0 & 22 & 3428.94 & 0.13 \\
0 & 23 & 3575.44 & 0.24 \\
0 & 24 & 3724.58 & 1.38 \\
0 & 25 & 3869.64 & 2.52 \\
\hline
1 & 14 & 2335.79 & 1.50 \\
1 & 15 & 2481.48 & 0.52 \\
1 & 16 & 2625.64 & 1.31 \\
1 & 17 & 2770.65 & 0.18 \\
1 & 18 & 2916.15 & 0.12 \\
1 & 19 & 3061.64 & 0.09 \\
1 & 20 & 3207.46 & 0.09 \\
1 & 21 & 3353.43 & 0.12 \\
1 & 22 & 3499.47 & 0.15 \\
1 & 23 & 3645.94 & 0.29 \\
1 & 24 & 3792.67 & 0.48 \\
1 & 25 & 3938.49 & 0.53 \\
\hline
2 & 17 & 2836.77 & 1.05 \\
2 & 18 & 2982.35 & 0.45 \\
2 & 19 & 3129.13 & 0.29 \\
2 & 20 & 3274.20 & 0.48 \\
2 & 21 & 3420.75 & 0.45 \\
2 & 22 & 3568.74 & 1.80 \\
2 & 23 & 3715.80 & 2.78 \\
2 & 24 & 3861.66 & 3.85 \\
\hline
\end{tabular}
\end{table}

\begin{table}
\caption{Observed frequency ratios of Kepler-93. These ratios were computed using the definitions of \citet{2003A&A...411..215R}.}
\begin{tabular}{cccc}
\hline
Ratio type & $n$ & Ratio & 68\% credible \\
\hline\hline
$r_{01}$ & 15 & 0.028 & 0.006 \\
$r_{01}$ & 16 & 0.031 & 0.011 \\
$r_{01}$ & 17 & 0.027 & 0.005 \\
$r_{01}$ & 18 & 0.023 & 0.001 \\
$r_{01}$ & 19 & 0.022 & 0.001 \\
$r_{01}$ & 20 & 0.021 & 0.001 \\
$r_{01}$ & 21 & 0.019 & 0.001 \\
$r_{01}$ & 22 & 0.018 & 0.001 \\
$r_{01}$ & 23 & 0.021 & 0.002 \\
$r_{01}$ & 24 & 0.032 & 0.008 \\
\hline
$r_{10}$ & 15 & 0.029 & 0.007 \\
$r_{10}$ & 16 & 0.030 & 0.010 \\
$r_{10}$ & 17 & 0.024 & 0.002 \\
$r_{10}$ & 18 & 0.022 & 0.001 \\
$r_{10}$ & 19 & 0.022 & 0.001 \\
$r_{10}$ & 20 & 0.020 & 0.001 \\
$r_{10}$ & 21 & 0.018 & 0.001 \\
$r_{10}$ & 22 & 0.018 & 0.001 \\
$r_{10}$ & 23 & 0.027 & 0.005 \\
$r_{10}$ & 24 & 0.031 & 0.010 \\
\hline
$r_{02}$ & 18 & 0.067 & 0.007 \\
$r_{02}$ & 19 & 0.067 & 0.003 \\
$r_{02}$ & 20 & 0.059 & 0.002 \\
$r_{02}$ & 21 & 0.062 & 0.003 \\
$r_{02}$ & 22 & 0.056 & 0.003 \\
$r_{02}$ & 23 & 0.046 & 0.012 \\
$r_{02}$ & 24 & 0.060 & 0.021 \\
$r_{02}$ & 25 & 0.055 & 0.032 \\
\hline
\end{tabular}
\end{table}

\clearpage

\section{Mode Identification}
\label{sec_appendix_mode_identification}
The $\epsilon_{nl}$ phases are defined in terms of the individual frequencies and their asymptotic formulation \citep{2016A&A...585A..63R}:
\begin{equation}
\epsilon_{nl} = \frac{\nu_{nl}}{\Delta} -n -\frac{l}{2},
\end{equation}
where $\nu_{nl}$ are the individual frequencies, $\Delta$ some arbitrary chosen reference large separation (e.g. Eq. (22) in \citet{2012A&A...539A..63R}), $n$ the radial order and $l$ the spherical degree.

\begin{figure}[h]
\centering
\includegraphics[scale=0.6]{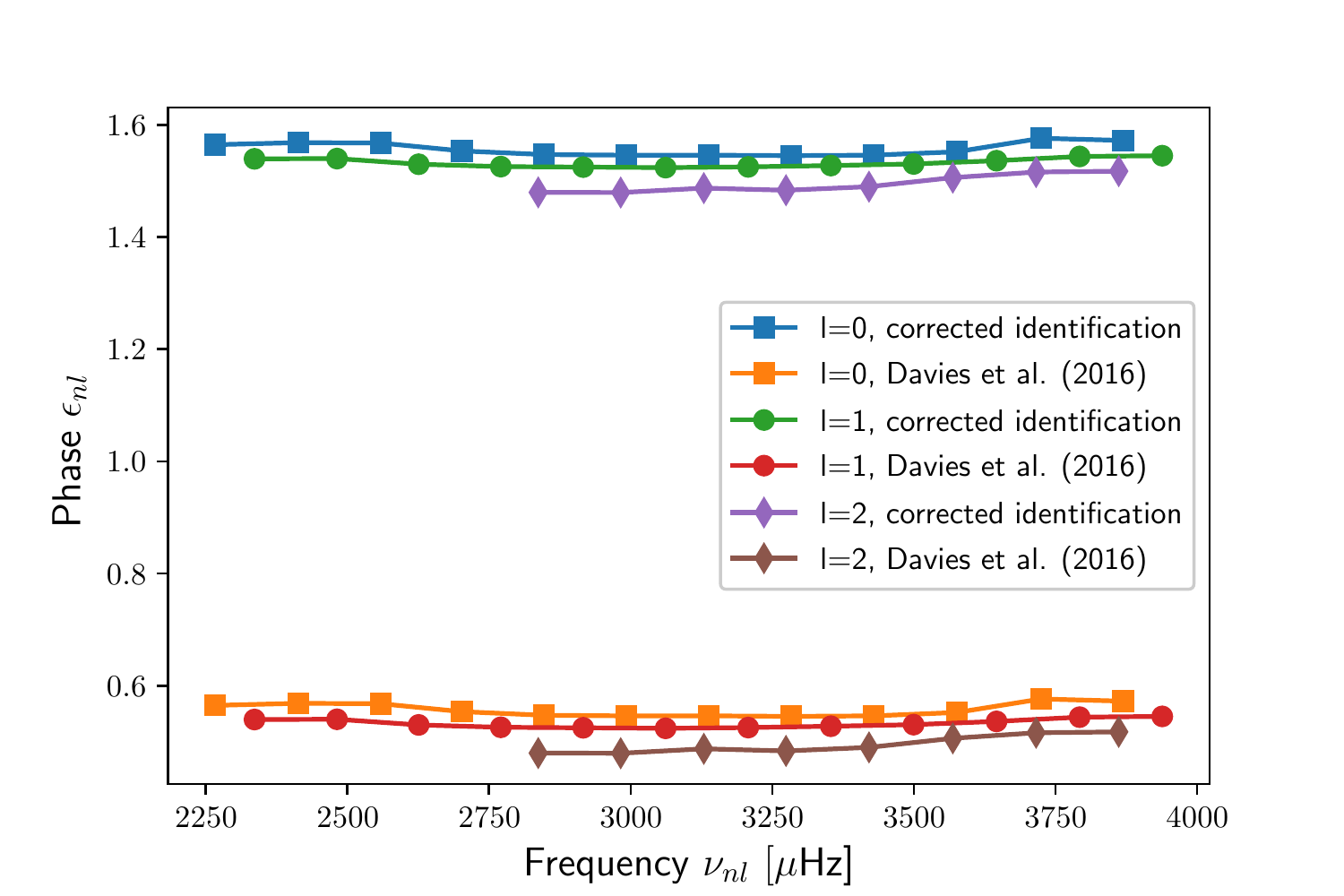}
\caption{Verification of the asymptotic behaviour of the $\epsilon_{nl}$ phases. The mode identification of \citet{2016MNRAS.456.2183D} gives phases below 1, what is unexpected. The corrected identification leads to results comparable to \citet{2016A&A...585A..63R}.}
\end{figure}

\clearpage

\onecolumn
\section{MCMC Corner Plot}
\label{sec_appendix_mcmc_corner_plot}
\begin{figure}[h]
\centering
\includegraphics[scale=0.4]{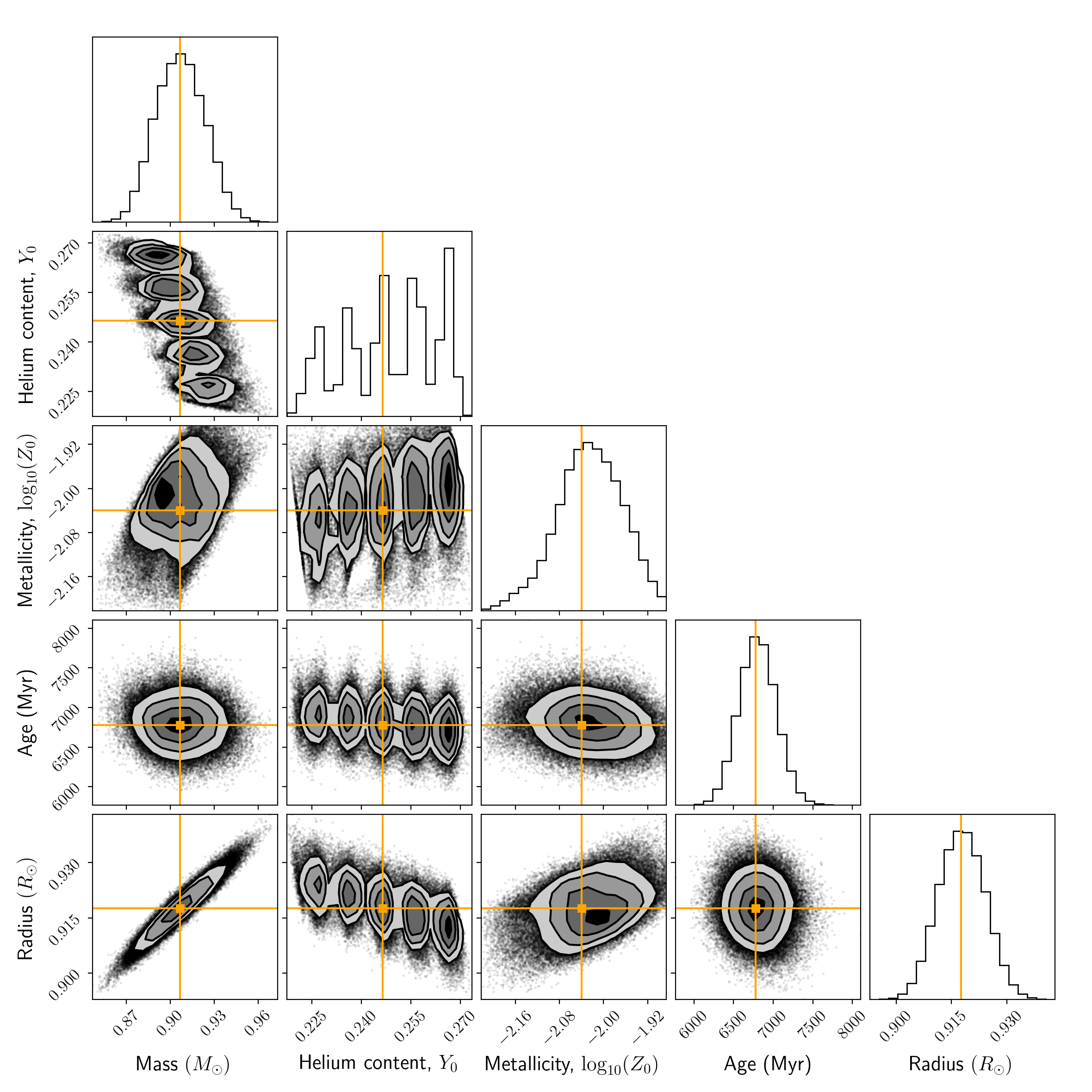}
\caption{Corner plot of the MCMC fitting the frequency ratios. The inverted mean density was part of the constraints. The optimal values are displayed in orange.}
\end{figure}

\end{document}